\documentclass[11pt]{article}
\usepackage{amssymb}
\usepackage{amsmath}
\usepackage{amscd}
\usepackage{latexsym}

\catcode`@=11
\renewcommand{\section}{\@startsection{section}{1}{0pt}{\medskipamount}
{\medskipamount}{\large\bf}}
\numberwithin{equation}{section}
\catcode`@=12
\def\a{\alpha}
\def\b{\beta}
\def\g{\gamma}

\def\de{\delta}
\def\ve{\varepsilon}
\def\h{\eta}
\def\th{\theta}
\def\vt{\vartheta}
\def\la{\lambda}
\def\m{\mu}
\def\n{\nu}

\def\s{\sigma}
\def\p{\phi}

\def\o{\omega}
\def\Om{\Omega}
\def\La{\Lambda}

\def\At{\widetilde A}
\def\Ft{\widetilde F}
\def\1{\bar 1}
\def\2{\bar 2}
\def\3{\bar 3}
\def\4{\bar 4}
\def\Ups{\Upsilon}

\def\lrc{\lrcorner}
\def\hra{\hookrightarrow}

\newcommand{\Ah}{\hat{A}}
\newcommand{\Qh}{\hat{Q}}
\newcommand{\Fh}{\hat{F}}

\newcommand{\yb}{\bar{y}}
\newcommand{\ab}{\bar{\alpha}}
\newcommand{\bb}{\bar{\beta}}

\newcommand{\zb}{\bar{z}}
\newcommand{\zeb}{\bar{\zeta}}

\newcommand{\ib}{\bar{i}}
\newcommand{\jb}{\bar{j}}
\newcommand{\kb}{\bar{k}}

\newcommand{\C}{\mathbb C}
\newcommand{\R}{\mathbb R}
\newcommand{\T}{\mathbb T}

\newcommand{\Zcal}{{\cal Z}}
\newcommand{\Acal}{{\cal A}}
\newcommand{\Ncal}{{\cal N}}
\newcommand{\Lcal}{{\cal L}}
\newcommand{\Fcal}{{\cal F}}
\newcommand{\Ecal}{{\cal E}}
\newcommand{\J}{{\cal J}}
\newcommand{\U}{{\cal U}}
\newcommand{\Tcal}{{\cal T}}
\newcommand{\Pcal}{{\cal P}}
\newcommand{\with}{{\qquad{\rm with}\qquad}}

\def\im{\mbox{i}}
\def\N2{$N{=}2$}
\def\pa{\mbox{$\partial$}}
\def\diff{\mbox{d}}
\def\tr{{\rm tr}}
\def\sfrac#1#2{{\textstyle\frac{#1}{#2}}}
\def\>{\rangle}
\def\<{\langle}
\def\+{\dagger}
\def\={\ =\ }
\def\und{\qquad\textrm{and}\qquad}
\def\and{\quad\textrm{and}\quad}

\begin{document}
\begin{titlepage}
\setcounter{page}{0}
\begin{flushright}
.
\end{flushright}

\vskip 2.5cm

\begin{center}

{\Large\bf
Hermitian-Yang-Mills equations and pseudo-holomorphic bundles\\[6pt]
on nearly K\"ahler and nearly Calabi-Yau twistor 6-manifolds}

\vspace{12mm}
{\large Alexander D. Popov}
\\[2mm]
\noindent {\em Bogoliubov Laboratory of Theoretical Physics, JINR\\
141980 Dubna, Moscow Region, Russia}
\\
{Email: {\tt popov@theor.jinr.ru}}
\vspace{12mm}

\begin{abstract}
\noindent
We consider the Hermitian-Yang-Mills (HYM) equations for gauge
potentials on a complex vector bundle $\Ecal$ over an almost
complex manifold $X^6$ which is the twistor space of an oriented
Riemannian manifold $M^4$. Each solution of the HYM equations
on such $X^6$ defines a pseudo-holomorphic structure on the
bundle $\Ecal$. It is shown that the pull-back to $X^6$ of any
anti-self-dual gauge field on $M^4$ is a solution of the HYM
equations on $X^6$. This correspondence allows us to introduce
new twistor actions for bosonic and supersymmetric Yang-Mills
theories. As examples of $X^6$ we consider homogeneous
nearly K\"ahler and nearly Calabi-Yau manifolds which are
twistor spaces of $S^4$, $\C P^2$ and $B_4$, $\C B_2$ (real 4-ball
and complex 2-ball), respectively. Various explicit examples of
solutions to the HYM equations on these spaces are provided.
Applications in flux compactifications of heterotic strings are
briefly discussed.
\end{abstract}
\end{center}

\end{titlepage}

\section{Introduction}

\noindent
In the recent paper~\cite{Lust}, the flux compactifications of type
IIA string theory of the form AdS$_4\times X^6$ with nearly K\"ahler
internal spaces $X^6=\ $Sp(2)/Sp(1)$\times$U(1) or SU(3)/U(1)$\times$U(1)
were considered (see also~\cite{Toma}) and it was found that the Kaluza-Klein
decoupling for the original AdS$_4$ vacua requires that the above-mentioned
internal spaces are substituted by the nearly Calabi-Yau spaces
Sp(1,1)/Sp(1)$\times$U(1), SU(1,2)/U(1)$\times$U(1) or by their compact
analogues obtained by quotienting out the internal manifolds by a discrete
group. In our paper, we describe various solutions of the
Hermitian-Yang-Mills equations on all these four coset spaces $X^6$ that
can further be used in heterotic string compactifications with non-trivial
background fluxes~\cite{FL}-\cite{Zoupanos}.

Since their discovery, more than ten years ago, tractable flux
compactifications in string theory have become a very active area of
research (see e.g.~\cite{Grana, DoKa, BKLS} for reviews and references).
This has been particularly explored in type IIB theory (see
e.g.~\cite{GKP}-\cite{LRS}), and some efforts have been devoted to
moduli-fixing problem in the case of type IIA compactifications (see
e.g.~\cite{LoMi, DKPZ, GL}) where metric fluxes can arise partially
from the T-duality of NS fluxes~\cite{GLMW, KSTT}. Compactifications in
the presence of fluxes can be described in the language of $G$-structures on
$d$-dimensional manifolds: SU(3) structure for dimension $d{=}6$, $G_2$
for $d{=}7$ and Spin(7) for $d{=}8$ (see e.g.~\cite{GMPW}-\cite{XuPhD}
and references therein). Note that 6-manifolds with SU(3) structure
(i.e. SU(3) holonomy of the spin connection with torsion proportional
to the $H$-field) can be described in terms of conditions on torsion
classes of these manifolds~\cite{CS}. Due to the
inclusion of the $H$-field in the geometry of the internal manifold
as torsion, we have to deal with non-K\"ahler and in some cases
non-complex manifolds.

Most research in flux compactifications was done in type II string theories
(see e.g.~\cite{Grana}-\cite{BKLS} and \cite{group0}-\cite{group2} for
more recent results). However, fluxes in heterotic string theory, which
play a prominent role in stringy model building, have been considered as
well (see e.g.~ \cite{FL, BLP, Zoupanos}, \cite{LM}-\cite{papa} and references
therein). Historically, heterotic flux compactifications have been known
for quite some time, starting with the works~\cite{SHW} in the mid-1980's.
In heterotic string compactifications one has the freedom to choose
a gauge bundle since the simple embedding of the spin connection into
the gauge connection is ruled out for compactifications with $dH\ne 0$. For
the torsionful background, the allowed gauge bundle is restricted by the
Hermitian-Yang-Mills equations~\cite{Don, UY} and by the Bianchi identity
for the $H$-field (anomaly cancellation). The existence of such vector
bundles on some non-K\"ahler {\it complex} 3-folds, their stability and the
procedure of solving the Hermitian-Yang-Mills equations were discussed e.g.
in~\cite{BB, Yau, Yaugroup, Fern}. Here we consider the procedure of solving
the Hermitian-Yang-Mills equations on the homogeneous nearly K\"ahler
spaces Sp(2)/Sp(1)$\times$U(1), SU(3)/U(1)$\times$U(1) and nearly Calabi-Yau
spaces Sp(1,1)/Sp(1)$\times$U(1), SU(1,2)/U(1)$\times$U(1) which may serve
as a local model for compact case obtained by quotienting out by a discrete
group.

The above-mentioned four manifolds are twistor spaces of the
four-dimensional sphere $S^4$, projective plane $\C P^2$ and balls
$B_4=\ $Sp(1,1)/Sp(1)$\times$Sp(1), $\C B_2=\ $SU(1,2)/S(U(1)$\times$U(2))
endowed with nonintegrable almost complex structure. Hence, complex
vector bundles over these twistor spaces can carry a pseudo-holomorphic
structure but not the holomorphic ones. That is why we begin our discussion
with the notion of the pseudo-holomorphic bundles~\cite{Bryant} and their
relations with the Hermitian-Yang-Mills equations. Then we consider the
twistor space $\Tcal (M^4)$ of an oriented Riemannian 4-manifold $M^4$
along with the canonical projection $\pi : \Tcal (M^4)\to M^4$ and a
complex vector bundle $E$ with an anti-self-dual connection $A$. We show that
any anti-self-dual gauge field $F=\diff A+A\wedge A$ on $M^4$ uplifted
to the gauge field $\hat F:=\pi^*F$ on $\Tcal (M^4)$ provides a solution
to the Hermitian-Yang-Mills equations on $\Tcal (M^4)$. This correspondence
allows us to introduce new twistor actions for bosonic and $\Ncal{=}4$
supersymmetric Yang-Mills theories.

Specializing to the cases $M^4=S^4$, $\C P^2$, $B_4$ and $\C B_2$, we
describe K\"ahler, nearly K\"ahler and nearly Calabi-Yau structures on
the twistor spaces $\Tcal (M^4)$ for these four cases. Various explicit
solutions of the Hermitian-Yang-Mills equations on $\Tcal (M^4)$ will be
written down and their applications in flux compactifications of heterotic
strings will be briefly discussed.

\bigskip

\section{Pseudo-holomorphic bundles and Hermitian-Yang-Mills equations}

\noindent
{\bf Notation}.
Let $X^{2n}$ be an oriented Riemannian $2n$-dimensional manifold and
$\{e^a\}$ a local orthonormal basis of $T^*X^{2n}$ , $a=1,...,2n$.
For $p$-forms on $X^{2n}$ we use the notation
\begin{equation}\label{2.1}
F_p=\sfrac{1}{p!}\,F_{a_1...a_p}\, e^{a_1...a_p}\with
e^{a_1...a_p}:=e^{a_1}\wedge ... \wedge e^{a_p}\ ,
\end{equation}
\begin{equation}\label{2.2}
(*F_p)_{a_1...a_{2n-p}}=\sfrac{1}{p!}\,\ve_{a_1...a_{2n-p}\, b_1...b_p}\,
F^{b_1...b_p}\quad\Leftrightarrow\quad
F_p\wedge *F_p= \sfrac{1}{p!}\, F_{a_1...a_p}\,F^{a_1...a_p}\,{vol}_{2n} \ ,
\end{equation}
where $*$ is the Hodge star operator and
${vol}_{2n}=e^1\wedge...\wedge e^{2n}$.
We also use notation~\cite{CS}
\begin{equation}\label{2.3}
(F_p\lrc\, S_{p+q})_{b_1...b_q}=\sfrac{1}{p!}\, (F_p)^{a_1...a_p}
(S_{p+q})_{a_1...a_p\,b_1...b_q}
\end{equation}
that exploits the underlying Riemannian metric
$\diff s^2 = \de_{ab}\, e^ae^b$
with the convention that $e^{12}\lrc\, e^{123}=e^3$ etc.

\smallskip

\noindent
{\bf Pseudo-holomorphic bundles}. Consider an oriented $2n$-dimensional
manifold $X^{2n}$ with an almost complex structure $\J$ and a complex
vector bundle $\Ecal$ over $X^{2n}$ endowed with a connection $\Acal$.
According to R.Bryant~\cite{Bryant}, a connection $\Acal$ on $\Ecal$ is said
to define a pseudo-holomorphic structure if it has curvature $\Fcal =
\diff\Acal +\Acal\wedge\Acal$ of type (1,1) with respect to (w.r.t.) $\J$,
i.e.
\begin{equation}\label{2.4}
\Fcal^{0,2}=0=\Fcal^{2,0}\ .
\end{equation}

One can endow the bundle $\Ecal$ with a Hermitian metric and choose $\Acal$
to be compatible with the Hermitian structure on $\Ecal$. If, in addition,
$\o$ is an almost Hermitian structure on $(X^{2n}, \J)$ and
\begin{equation}\label{2.5}
\o\,\lrc\,\Fcal = \im\, \la\, \rm{Id}_{\Ecal}
\end{equation}
with $\la\in\R$, the connection $\Acal$ is said to be
(pseudo-)Hermitian-Yang-Mills~\cite{Bryant}. We shall consider (\ref{2.5}) with
$\la =0$, i.e. assume $c_1(\Ecal )=0$ since for a bundle with field
strength $\Fcal$ of non-zero degree one can obtain a zero-degree bundle
by considering $\widetilde\Fcal =\Fcal - \sfrac{1}{k}(\tr\Fcal)\cdot
{\bf 1}_k$, where $k=\,$rank$\,\Ecal$.

\smallskip

\noindent
{\bf Hermitian-Yang-Mills equations}. The Hermitian-Yang-Mills (HYM)
equations\footnote{We omit the prefix `pseudo' for conformity with
the literature on string compactifications.} read
\begin{equation}\label{2.6}
\Fcal^{0,2}=0\und\o\,\lrc\,\Fcal =0\ .
\end{equation}
In the special case of an almost complex 4-manifold $X^4$ with a metric
$g$ they coincide with the anti-self-dual Yang-Mills (ASDYM) equations
\begin{equation}\label{2.7}
*\Fcal =-\Fcal\ ,
\end{equation}
where $*$ is the Hodge operator. Note that (\ref{2.7}) is valid on manifolds
$(M^4, g)$ which are not necessarily almost complex manifolds.

Recall that there are various generalizations of the first order
ASDYM equations (\ref{2.7}) to higher dimensions~\cite{CDFN}-\cite{Tian}
with some explicit solutions (see e.g.~\cite{FN}).
In particular, in $d{=}2n{=}6$ one can consider the equations~\cite{Tian}
\begin{equation}\label{2.8}
*\Fcal =-\o\wedge\Fcal\ ,
\end{equation}
where $\o$ is a two-form. Differentiating (\ref{2.8}), we obtain
\begin{equation}\label{2.9}
\diff (*\Fcal)+\Acal\wedge*\Fcal -*\Fcal\wedge\Acal+*H\wedge\Fcal =0\ ,
\end{equation}
where the 3-form $H$ is defined by the formula
\begin{equation}\label{2.10}
H:=*\diff\o\ .
\end{equation}
Equations (\ref{2.9}) differ from the standard Yang-Mills equations
by the last term with a 3-form $H$ which can be identified with a totally
antisymmetric torsion. These equations naturally appear in string theory.

For manifolds $X^6$ with an almost complex structure $\J$ the equations
(\ref{2.8}) can be rewritten in the form (\ref{2.6}) with an almost
Hermitian structure $\o$. To each solution $\Acal$ of the HYM equation
(\ref{2.6}) there corresponds a pseudo-holomorphic structure on the
vector bundle $\Ecal$ over $X^6$. In the case of integrable almost
complex structure $\J$ the first equation in (\ref{2.6}) defines a
holomorphic structure on $\Ecal$ and the second equation in (\ref{2.6})
is the requirement of (semi)stability of the bundle $\Ecal$~\cite{Don,UY}.
Thus, for complex manifolds the HYM connections $\Acal$ (solutions
to (\ref{2.6})) describe (semi)stable holomorphic bundles $\Ecal$.
It would be interesting to generalize the notion of stability to
pseudo-holomorphic bundles $\Ecal$ and to learn whether the second
equation in (\ref{2.6}) is also equivalent to an expected stability
of $\Ecal$.

\bigskip

\section{Twistor correspondence and pseudo-holomorphic bundles}
\smallskip

\noindent
{\bf Twistor space of $M^4$}. Let us consider an oriented real
four-manifold\footnote{It is not necessary that this manifold
is almost complex. For instance, there is not any almost complex
structure on the four-sphere $S^4$.} with a Riemannian metric
$g$ and the principal bundle $P(M^4, SO(4))$ of orthonormal frames
over $M^4$. The twistor space $\Tcal (M^4)$ of $M^4$ can be defined
as an associated bundle~\cite{AHS}
\begin{equation}\label{3.1}
\Tcal (M^4) = P\times_{SO(4)} SO(4)/U(2)
\end{equation}
with the canonical projection
\begin{equation}\label{3.2}
\pi : \Tcal (M^4) \to M^4\ .
\end{equation}
The fibres of this bundle are two-spheres $S^2_x\cong\,$SO(4)/U(2)
which parametrize almost complex structures on the tangent spaces
$T_x M^4$. As a real manifold, $\Tcal (M^4)$ has dimension six.

Another (equivalent) definition of $\Tcal (M^4)$ is obtained by
considering the vector bundle $\La^2T^*M^4$ of two-forms on $M^4$.
Using the Hodge operator (\ref{2.2}), one can split $\La^2T^*M^4$
into the direct sum $\La^2T^*M^4=\La_+^2T^*M^4\oplus \La_-^2T^*M^4$
of the subbundles of self-dual and anti-self-dual two-forms on $M^4$.
Then the twistor space $\Tcal (M^4)$ of $M^4$ can be defined as the
unit sphere bundle
\begin{equation}\label{3.3}
\Tcal (M^4)= S_1(\La_+^2T^*M^4)
\end{equation}
in the vector bundle $\La_+^2T^*M^4$.

Note that while a manifold $M^4$ admits in general no almost complex
structure, its twistor space $\Tcal (M^4)$ can always be equipped with
two natural almost complex structures. The first, $\J =\J_+$,
introduced in~\cite{AHS}, is integrable if and only if the Weyl tensor
of $g$ on $M^4$ is anti-self-dual, while the second $\J =\J_-$,
introduced in~\cite{ES}, is never integrable. In fact, $\J_+$ and $\J_-$
differ only on $S^2_x\cong\C P^1\hra\Tcal (M^4)$ ($\J_+|_{\C P^1}
= -\J_-|_{\C P^1}$) and coincide on $T_xM^4$. Twistor spaces
$\Tcal (M^4)$ with an almost complex structure $\J$ can be considered
as a particular case of almost complex manifolds $X^6$ discussed in
section 2 in the context of the pseudo-holomorphic bundles and the
HYM equations.

\smallskip

\noindent
{\bf Pull-back of complex vector bundles from $M^4$ to $\Tcal (M^4)$}.
Let $E$ be a rank $k$ complex vector bundle over $M^4$ and $A$ a
connection one-form (gauge potential) on $E$ with the curvature
$F=\diff A+A\wedge A$ (gauge field). Suppose that the gauge field $F$
satisfies the ASDYM equations (\ref{2.7}). Bundles $E$ with such
a connection $A$ are called anti-self-dual~\cite{AHS}. Using the
projection (\ref{3.2}), we pull $E$ back to a bundle $\hat E:=\pi^*E$
over $\Tcal (M^4)$. In accordance with the definition of a pull-back,
the connection $\Ah :=\pi^* A$ on $\hat E$ is flat along the fibres
$\C P^1_x$ of the bundle (\ref{3.2}) and we can set the components of
$\Ah$ along the fibres equal to zero. Thus, restrictions of the
smooth vector bundle $\hat E$ to fibres $\C P^1_x$ of projection $\pi$
are holomorphically trivial for each $x\in M^4$.

It was shown in~\cite{AHS} that if the Weyl tensor of $(M^4, g)$
is anti-self-dual then the almost complex structure $\J=\J_+$ on the
twistor space $\Tcal (M^4)$ is integrable and $\Tcal (M^4)$ inherits
the structure of a complex analytic 3-manifold. Furthermore, it was
proven that an anti-self-dual bundle $E$ over anti-self-dual $M^4$
lifts to a holomorphic bundle $\hat E$ over complex $\Tcal (M^4)$
defined by the equation $\Fh^{0,2}=0$, where $\Fh :=\pi^*F$ is the
pull-back to $\hat E$ of an anti-self-dual (ASD) gauge field $F$ on $E$.
In~\cite{AHS} it was also mentioned in a remark that one can introduce
a Hermitian metric on $\Tcal (M^4)$ such that $\Fh$ will be orthogonal
to the Hermitian form. However, the HYM equations were introduced
later~\cite{Don, UY} in a different context.

\smallskip

\noindent
{\bf Generalized twistor correspondence}. The essence of the canonical
twistor approach is to establish a correspondence between four-dimensional
space-time $M^4$ and {\it complex} twistor space $\Tcal (M^4)$ of $M^4$.
Using this correspondence, one transfers data given on $M^4$ to data on
$\Tcal (M^4)$ and vice versa. In twistor theory one considers
{\it holomorphic} objects  $h$ on $\Tcal (M^4)$ (\v Cech cohomology
classes, holomorphic vector bundles etc.) and transforms them to objects
$f$ on $M^4$ which are constrained by some differential equations
\cite{Penrose, Ward77, AHS, Wells}. Thus, the main idea
of twistor theory is to encode solutions of some differential equations
on $M^4$ in holomorphic data on the complex twistor space $\Tcal (M^4)$ of
$M^4$. In particular, solutions of the ASDYM equations on manifolds
$M^4$ with the ASD Weyl tensor correspond to holomorphic vector bundles
$\hat E$ over $\Tcal (M^4)$. However, in Donaldson theory~\cite{DK} one
considers the ASDYM equations on manifolds $M^4$ whose Weyl tensor is not
restricted and it is desirable to have a twistor description of this generic
case.\footnote{This desire is supported by ideas of the theory of harmonic
maps where never integrable almost complex structure $\J_-$ on twistor
spaces play a key role~\cite{ES, Salamon}. For a recent review of this
subject see e.g.~\cite{Sergeev}.}.

In~\cite{Popov} it has been shown that the vortex equations on a compact
Riemann surface $\Sigma$ are equivalent to the ASDYM equations on
$\Sigma\times\C P^1$ and to the Hermitian-Yang-Mills equations on the
twistor space $\Tcal (M^4)$ of $M^4=\Sigma\times\C P^1$. In the general
case, the manifold $M^4$ is not anti-self-dual and an almost complex
structure on $\Tcal (M^4)$ is not integrable.\footnote{In a special case,
when the twistor geometry becomes integrable (holomorphic), the vortex
equations on $\Sigma$ appear as a commutator of two auxiliary linear
differential operators with a `spectral' parameter, i.e. become integrable.}
Here, we show that this generalized twistor correspondence holds for the
case of an arbitrary oriented 4-manifold $M^4$. Namely, we describe a
correspondence between Hermitian vector bundles $E$ with ASD connections
$A$ on an oriented 4-manifold $M^4$ and pseudo-holomorphic vector bundles
$\hat E$ over an almost complex twistor space $\Tcal (M^4)$.

\smallskip

\noindent
{\bf Almost complex structure on $\Tcal (M^4)$}. We fix an open
subset\footnote{This subset may coinside with a point $x\in M^4$.}
$\U$ of $M^4$ with coordinates $\{x^{\m}\}$, $\m =1,...,4$, and an open
subset $U=\C P^1\setminus\{\infty\}$ of $\C P^1$ with a local complex coordinate
$\zeta$. Then $\U\times U$ is an open subset of $\Tcal (M^4)$. Note
that over $\U$ there exists a section $J=(J^\n_\m )$ of the bundle
(\ref{3.2}) (a local almost complex structure) and this allows to
introduce forms of type $(p,q)$ w.r.t. $J$. Globally such an almost
complex structure $J$ on $M^4$ may not exist but this is not necessary
for all twistor constructions.

Let $\{\vt^\m\}$ represents some orthonormal coframe on $\U\subset M^4$,
i.e. $\diff s^2=\de_{\m\n}\vt^\m\vt^\n$. Using the canonical form of
a local almost complex structure $J$, we introduce forms
\begin{equation}\label{3.4}
\th^1:=\vt^1+\im\vt^2\ ,\quad\th^2:=\vt^3+\im\vt^4\ ,\quad\th^{\1}:=\vt^1-
\im\vt^2\und\th^{\2}:=\vt^3-\im\vt^4\ ,
\end{equation}
which provide a local basis of orthonormal (1,0)-forms w.r.t. $J$.
Then one can introduce forms
\begin{equation}\label{3.5}
\o^1:=\frac{1}{(1{+}\zeta\zeb)^\frac{1}{2}}(\th^1{-}\zeta\th^{\2})\ ,
\quad
\o^2:=\frac{1}{(1{+}\zeta\zeb)^\frac{1}{2}}(\th^2{+}\zeta\th^{\1})
\and
\o^3:=\frac{1}{1{+}\zeta\zeb}(\diff\zeta{-}\Gamma_+^iL_i^\zeta)\ ,
\end{equation}
which may serve as the definition of an almost complex structure $\J$
on $\Tcal (M^4)$ such that
\begin{equation}\label{3.6}
\J\,\o^i =\im\,\o^i\quad{\rm for}\quad i=1,2,3\ .
\end{equation}
Here $\Gamma_+=(\Gamma_+^i)$ is the self-dual part of the Levi-Civita
connection on $M^4$ and $L_i^{\zeta}$ are holomorphic components of
vector fields $L_i=L_i^{\zeta}\pa_{\zeta}{+}L_i^{\zeb}\pa_{\zeb}$ on
fibres $\C P^1_x\hra \Tcal (M^4)$ which give a realization of the generators
of the group SU(2) acting on $\C P^1{=}$SU(2)/U(1). One can take
e.g. $L_1-\im L_2=-2\im$, $L_1+\im L_2=-2\im\zeta^2$ and $L_3=-2\im\zeta$.
Note that the forms (\ref{3.5}) extend to a basis of globally defined forms
on $\Tcal (M^4)$ of type (1,0) w.r.t. $\J$. That is why our discussion does
not depend on the choice of local coordinates, forms etc.

\smallskip

\noindent
{\bf From ASDYM on $M^4$ to HYM on $\Tcal (M^4)$}. For the curvature
$F=\diff A + A\wedge A$ of the vector bundle $E\to M^4$ we have
\begin{equation}\label{3.7}
F=\sfrac{1}{2}\,(F+*F) + \sfrac{1}{2}\,(F-*F)=:F^++F^-\ ,
\end{equation}
where in the basis (\ref{3.4}) of local forms
\begin{equation}\label{3.8}
F^+=\sfrac{1}{2}\,(F_{1\1}+F_{2\2})(\th^{1\1}+\th^{2\2})+F_{12}\th^{12}
+F_{\1\2}\th^{\1\2}\ ,
\end{equation}
\begin{equation}\label{3.9}
F^-=\sfrac{1}{2}\,(F_{1\1}-F_{2\2})(\th^{1\1}-\th^{2\2})+F_{1\2}\th^{1\2}
+F_{2\1}\th^{2\1}\ ,
\end{equation}
with $\th^{1\1}:=\th^{1}\wedge\th^{\1},\ \th^{12}:=\th^{1}\wedge\th^{2}$
etc. Here $F^+$ and $F^-$ are self-dual and anti-self-dual parts of the
curvature $F$, respectively.

For the pull-back $\Fh^{\pm}:=\pi^*F^{\pm}$ of the two-forms (\ref{3.8})
and (\ref{3.9}) on $M^4$ to $\Tcal (M^4)$ we obtain
\begin{equation}\label{3.10}
\Fh^+=\sfrac{1}{2}\,(\Fh_{1\1}+\Fh_{2\2})(\o^{1\1}+\o^{2\2})+
\Fh_{12}\o^{12}+\Fh_{\1\2}\o^{\1\2}\ ,
\end{equation}
\begin{equation}\label{3.11}
\Fh^-=\sfrac{1}{2}\,(\Fh_{1\1}-\Fh_{2\2})(\o^{1\1}-\o^{2\2})+
\Fh_{1\2}\o^{1\2}+\Fh_{2\1}\o^{2\1}\ ,
\end{equation}
where $\o^{1\1}:=\o^{1}\wedge\o^{\1},\ \o^{12}:=\o^{1}\wedge\o^{2}$
etc. Note that
\begin{eqnarray}
\o^{12}=\frac{1}{1{+}\zeta\zeb}\,[\th^{12}+\zeta (\th^{1\1}+\th^{2\2})
+\zeta^2\th^{\1\2}]\ ,\quad \o^{1\2}=\th^{1\2}\ ,\nonumber\\
\o^{\1\2}=\frac{1}{1{+}\zeta\zeb}\,[\th^{\1\2}-\zeb (\th^{1\1}+\th^{2\2})
+\zeb^2\th^{12}]\ ,\quad \o^{2\1}=\th^{2\1}\ ,\nonumber\\
\o^{1\1}{+}\o^{2\2}=\frac{1}{1{+}\zeta\zeb}\,[(1{-}\zeta\zeb)
(\th^{1\1}{+}\th^{2\2}){+}2\zeta\th^{\1\2}{-}2\zeb\th^{12}]\ ,\quad
\o^{1\1}{-}\o^{2\2}=\th^{1\1}{-}\th^{2\2}\ ,\label{3.12}
\end{eqnarray}
as one can easily derive from (\ref{3.5}). Also we have
\begin{eqnarray}
\Fh_{12}=\frac{1}{1{+}\zeta\zeb}\,[F_{12}+\zeb (F_{1\1}+F_{2\2})
+\zeb^2F_{\1\2}]\ ,\quad \Fh_{1\2}=F_{1\2}\ ,\nonumber\\
\Fh_{\1\2}=\frac{1}{1{+}\zeta\zeb}\,[F_{\1\2}-\zeta (F_{1\1}+F_{2\2})
+\zeta^2F_{12}]\ ,\quad \Fh_{2\1}=F_{2\1}\ ,\nonumber\\
\Fh_{1\1}{+}\Fh_{2\2}=\frac{1}{1{+}\zeta\zeb}\,[(1{-}\zeta\zeb)
(F_{1\1}{+}F_{2\2}){-}2\zeta F_{12}{+}2\zeb F_{\1\2}]\ ,\quad
\Fh_{1\1}{-}\Fh_{2\2}=F_{1\1}{-}F_{2\2}\ ,\label{3.13}
\end{eqnarray}
and by construction
\begin{equation}\label{3.14}
\Fh_{i\3}=\Fh_{i3}=0\und\rm{h.c.}
\end{equation}
for $i=1,2,3$.

Using (\ref{3.5}), we can introduce on $\Tcal (M^4)$ an almost Hermitian
form
\begin{equation}\label{3.15}
\o =\sfrac{\im}{2}\,\left(\o^1\wedge\o^{\1}+\o^2\wedge\o^{\2}+
\ve\o^3\wedge\o^{\3}\right)\ ,
\end{equation}
where $\ve=\pm 1$.\footnote{Note that the metric for $\ve =-1$ will
have Hermitian signature (2,1). Later we shall see that $\ve =-1$ can
be a proper choice for manifolds $M^4$ of negative scalar curvature.}
Then for $\Fh =\Fh^-$ from (\ref{3.11}) it follows that
\begin{equation}\label{3.16}
\Fh^{0,2}=0
\end{equation}
and
\begin{equation}\label{3.17}
\o\,\lrc\,\Fh =0\ .
\end{equation}
Thus, anti-self-dual gauge fields $F=F^-$ on the vector bundle $E\to M^4$
are pulled back to the gauge fields $\Fh$ on the vector bundle $\hat E$
over the twistor space $\Tcal (M^4)$ which satisfy the Hermitian-Yang-Mills
equations (\ref{3.16}), (\ref{3.17}) on $\Tcal (M^4)$ without demanding
integrability of an almost complex structure (\ref{3.6}). In its turn,
such gauge fields $\Fh$ define a pseudo-holomorphic structure on the
vector bundle $\hat E$ which is holomorphically trivial on $\C P^1_x\hra
\Tcal (M^4)$ for each $x\in M^4$. Conversely, any such pseudo-holomorphic
bundle $\hat E\to\Tcal (M^4)$ corresponds to a solution $A$ of the ASDYM
equations on $M^4$.

\bigskip

\section{K\"ahler geometry on twistor spaces of $S^4$ and $B_4$}

Here, as $M^4$ we consider the four-sphere $S^4$ with a metric $g$ of
constant positive curvature and the open four-ball $B_4$ with a metric
$g$ of constant negative curvature. In the next section~5 we shall
consider the projective plane $\C P^2$ with the Fubini-Study metric
and the complex two-ball $\C B_2$ with the metric of constant negative
holomorphic sectional curvature. All these spaces $M^4$ are homogeneous
manifolds (coset spaces) as well as their twistor spaces $\Tcal (M^4)$.
Although the geometry of these spaces is well-known, we describe it
here by using local coordinates for fixing our notation. We also need
this for self-consistency and further applications.

\smallskip

\noindent
{\bf Manifolds $S^4$ and $B_4$ as coset spaces}. Let us consider the
group Sp(2) as a subgroup of SU(4) and the group Sp(1,1) as a
subgroup of SU(2,2) defined as 4$\times$4 matrices $Q$ such that
\begin{equation}\label{4.1}
Q^\+\eta\,Q = Q\eta\, Q^\+=\eta\with
\eta =\rm{diag}({\bf 1}_2, \ve{\bf 1}_2)\ ,
\end{equation}
where $\ve =1$ for Sp(2)$\,\subset\,$SU(4) and $\ve =-1$ for
Sp(1,1)$\,\subset\,$SU(2,2). We consider $S^4$ and $B_4$ as coset spaces
\begin{equation}\label{4.2}
S^4={\rm Sp}(2)/{\rm Sp}(1)\times {\rm Sp}(1)\und
B_4={\rm Sp}(1,1)/{\rm Sp}(1)\times {\rm Sp}(1)
\end{equation}
of positive and negative scalar curvature, respectively. Then one can
consider Sp(2) fibred over $S^4$ and Sp(1,1) fibred over $B_4$ as
principal bundles
\begin{equation}\label{4.3}
{\rm Sp}(2)\to S^4\und
{\rm Sp}(1,1)\to B_4\ ,
\end{equation}
both with the structure group Sp(1)$\times$Sp(1).

Let us consider local sections of the fibrations (\ref{4.3}) which are
given by 4$\times$4 matrices
\begin{equation}\label{4.4}
Q:= f^{-\frac{1}{2}}\begin{pmatrix}{\bf 1}_2 &-\ve x\\x^\+ &
{\bf 1}_2\end{pmatrix}
\und
Q^{-1} = \eta Q^\+\eta =f^{-\frac{1}{2}}\begin{pmatrix}{\bf 1}_2 &\ve x\\
-x^\+ & {\bf 1}_2\end{pmatrix}\ ,
\end{equation}
where
\begin{equation}\label{4.5}
x=x^\m\tau_\m\ ,\quad x^\+=x^\m\tau_\m^\+\ ,\quad
f:=1+\ve\,x^\+ x= 1+\ve\,r^2=1+\ve\de_{\m\n}x^\m x^\n\ ,
\end{equation}
and matrices
\begin{equation}\label{4.6}
(\tau_\m )= (-\im\s_i, {\bf 1}_2)\und
(\tau_\m^\+ )= (\im\s_i, {\bf 1}_2)
\end{equation}
obey
\begin{eqnarray}\nonumber
\tau_\m^\+ \tau_\n = \de_{\m\n}\cdot {\bf 1}_2 +
\eta_{\m\n}^i\,\im\,\s_i=:
\de_{\m\n}\cdot {\bf 1}_2 + \eta_{\m\n}\ ,\quad
\{\eta_{\m\n}^i\}=\{\ve^i_{jk}, \m{=}j, \n{=}k; \ \de_j^i,
\m{=}j, \n{=}4\}\ ,\\ \label{4.7}
\tau_\m\tau_\n^\+ = \de_{\m\n}\cdot {\bf 1}_2 + \bar\eta_{\m\n}^i\,
\im\,\s_i=: \de_{\m\n}\cdot {\bf 1}_2 + \bar\eta_{\m\n}\ ,\quad
\{\bar\eta_{\m\n}^i\}=\{\ve^i_{jk}, \m{=}j, \n{=}k; \ -\de_j^i,
\m{=}j, \n{=}4\}\ .
\end{eqnarray}
Here $\{x^\m\}$ are local coordinates on $\U\subset S^4$ or $B_4$. Note that
we will consistently combine formulae for both these spaces with the help
of $\ve =\pm 1$. Matrices (\ref{4.4}) are representative elements for
cosets (\ref{4.2}) encoding information about their differential geometry.

\smallskip

\noindent
{\bf (Anti-)self-dual gauge fields}. For $M^4=S^4$ or $B_4$, let us consider
a flat connection $\Acal$ on the trivial vector bundle $M^4\times\C^4\to M^4$
given by the one-form
\begin{equation}\label{4.8}
\Acal = Q^{-1}\diff Q =:
\begin{pmatrix}A^- &-\ve\p\\ \p^\+ & A^+\end{pmatrix} \ ,
\end{equation}
where from (\ref{4.4}) we obtain
\begin{equation}\label{4.9}
A^-= \sfrac{\ve}{f}\bar\h_{\m\n}x^\m\diff x^\n =:
\begin{pmatrix}\a_- &-\bar\b_-\\ \b_- & -\a_-\end{pmatrix}\in su(2) \ ,
\end{equation}
\begin{equation}\label{4.10}
A^+= \sfrac{\ve}{f}\h_{\m\n}x^\m\diff x^\n =:
\begin{pmatrix}\a_+ &-\bar\b_+\\ \b_+ & -\a_+\end{pmatrix}\in su(2) \ ,
\end{equation}
\begin{equation}\label{4.11}
\p = \frac{1}{f}\diff x = -\frac{\im}{f}
\begin{pmatrix}\diff x^3{+}\im\diff x^4&  \diff x^1{-}\im\diff x^2\\
\diff x^1{+}\im\diff x^2&  -(\diff x^3{-}\im\diff x^4)\end{pmatrix}
=-\frac{\im}{f}\begin{pmatrix}\diff z&\diff\bar y\\ \diff y&-\diff\bar z
\end{pmatrix}=:-\frac{\im}{2\Lambda}\begin{pmatrix}\th^2&\th^{\1}\\
\th^1&-\th^{\2}\end{pmatrix}\ ,
\end{equation}
with
\begin{equation}\label{4.12}
\a_-= \sfrac{\ve}{2f}(\yb\,\diff y + \zb\,\diff z - y\,\diff \yb -
z\,\diff \zb)\ ,\quad \b_-= \sfrac{\ve}{f}(y\,\diff z -z \,\diff y)\ ,
\end{equation}
\begin{equation}\label{4.13}
\a_+= \sfrac{\ve}{2f}(\yb\,\diff y + z\,\diff \zb - y\,\diff \yb -
\zb\,\diff z)\ ,\quad \b_+= \sfrac{\ve}{f}(y\,\diff \zb -\zb \,\diff y)\ ,
\end{equation}
\begin{equation}\label{4.14}
\th^1:=\frac{2\La\diff y}{1+\ve r^2}\ ,\quad
\th^2:=\frac{2\La\diff z}{1+\ve r^2}\und
\th^{\1}:=\frac{2\La\diff \yb}{1+\ve r^2}\ ,\quad
\th^{\2}:=\frac{2\La\diff \zb}{1+\ve r^2}\ .
\end{equation}
Here, the bar denotes complex conjugation. Note that the real parameter
$\La$ can be identified with the ``radius'' of $M^4=S^4$ or $B_4$.

The connection (\ref{4.8}) is flat, i.e.
\begin{equation}\label{4.15}
\Fcal = \diff\Acal + \Acal\wedge\Acal = \begin{pmatrix}
F^--\ve\,\p\wedge\p^\+&-\ve (\diff\p + \p\wedge A^+ + A^-\wedge\p)\\
\diff\p^\+ + A^+\wedge\p^\+ + \p^\+\wedge A^-& F^+-\ve\,\p^\+\wedge\p
\end{pmatrix} =0\ ,
\end{equation}
where $F^\pm = \diff A^\pm + A^\pm\wedge A^\pm$. From (\ref{4.15})
we get
\begin{equation}\label{4.16}
F^- = \ve\,\p\wedge\p^\+ = - \frac{\ve}{4\La^2}
\begin{pmatrix} \th^1\wedge\th^{\1} - \th^2\wedge\th^{\2}& 2\th^{\1}
\wedge\th^2\\
 -2\th^1\wedge\th^{\2}&-\th^1\wedge\th^{\1} + \th^2\wedge\th^{\2}
\end{pmatrix}\ ,
\end{equation}
\begin{equation}\label{4.17}
F^+ = \ve\,\p^\+\wedge\p = - \frac{\ve}{4\La^2}
\begin{pmatrix} \th^1\wedge\th^{\1} + \th^2\wedge\th^{\2}&
2\th^{\1}\wedge\th^{\2}\\
 -2\th^1\wedge\th^{2}&-\th^1\wedge\th^{\1} - \th^2\wedge\th^{\2}
\end{pmatrix}\ .
\end{equation}
One can easily see that $*F^{\pm}=\pm F^{\pm}$, i.e. $F^+$ and $F^-$
are self-dual (SD) and anti-self-dual (ASD) gauge fields on a rank-2
complex vector bundle $E\to M^4$,
respectively. They can be identified with SD and ASD parts of the
Riemann tensor of the metric $\diff s^2=\th^1\th^{\1}+ \th^2\th^{\2}$.

\smallskip

\noindent
{\bf Twistor manifolds of $S^4$ and $B_4$ as coset spaces}. Let us
consider the Hopf bundle
\begin{equation}\label{4.18}
S^3\to S^2
\end{equation}
over the Riemann sphere $S^2\cong\C P^1$ and the one-monopole
connection $a$ on the bundle (\ref{4.18}) having in the local
coordinate $\zeta\in\C P^1$ the form
\begin{equation}\label{4.19}
a=\frac{1}{2(1+\zeta\bar\zeta)}\, (\bar\zeta\,\diff\zeta -\zeta\,
\diff\bar\zeta)\ .
\end{equation}
Consider a local section of the bundle (\ref{4.18}) given by the
matrix
\begin{equation}\label{4.20}
g=\frac{1}{(1+\zeta\bar\zeta)^{\frac{1}{2}}}\,
\begin{pmatrix}1&-\bar\zeta\\ \zeta & 1\end{pmatrix}
\in {\rm SU}(2)\cong S^3
\end{equation}
and introduce the $su(2)$-valued one-form (flat connection)
\begin{equation}\label{4.21}
g^{-1}\diff g=:
\begin{pmatrix}a&-\frac{1}{2R}\th^{\3}\\
\frac{1}{2R}\th^{3} & -a\end{pmatrix}\ ,
\end{equation}
where
\begin{equation}\label{4.22}
\th^3=\frac{2R\diff\zeta}{1+\zeta\bar\zeta}\und
\th^{\3}=\frac{2R\diff\bar\zeta}{1+\zeta\bar\zeta}
\end{equation}
are the forms of type (1,0) and (0,1) on $\C P^1$, $a$ is the
one-monopole gauge potential (\ref{4.19}) and $R$ is the radius
of the Riemann sphere $\C P^1$ with the metric
\begin{equation}\label{4.23}
\diff s^2_{\C P^1}=\th^3\th^{\3}=
\frac{4R^2\diff\zeta\,\diff\bar\zeta}{(1+\zeta\bar\zeta)^2}\ .
\end{equation}
The K\"ahler form on $\C P^1$ is
\begin{equation}\label{4.24}
\o_{\C P^1}=\sfrac{\im}{2}\,\th^3\wedge\th^{\3}\ .
\end{equation}

Let us introduce 4$\times$4 matrices
\begin{equation}\label{4.25}
G=\begin{pmatrix}{\bf 1}_2&0\\0&g\end{pmatrix}\und\hat Q=QG\in
\left\{\begin{array}{l}{\rm Sp}(2)\subset{\rm SU}(4)
\quad {\rm for}\quad \ve =1\\
{\rm Sp}(1,1)\subset{\rm SU}(2,2)
\quad {\rm for}\quad \ve =-1\end{array}\right.
\end{equation}
where $Q$ and $g$ are given in (\ref{4.4}) and (\ref{4.20}).
The matrix $\hat Q$ is a local section of the bundle
\begin{equation}\label{4.26}
{\rm Sp}(2) \to {\rm Sp}(2)/{\rm Sp}(1){\times}{\rm U}(1)
=:\Tcal (S^4)
\end{equation}
or
\begin{equation}\label{4.27}
{\rm Sp}(1,1) \to {\rm Sp}(1,1)/{\rm Sp}(1){\times}{\rm U}(1)
=:\Tcal (B_4)\ ,
\end{equation}
depending on the choice $\ve =1$ or $\ve =-1$. In (\ref{4.26}) and
(\ref{4.27}) the twistor spaces $\Tcal (S^4)$ and $\Tcal (B_4)$ appear
as coset spaces and the matrices $\hat Q(\ve=\pm 1)$ from (\ref{4.25})
are representatives for the cosets $\Tcal (S^4)$ and $\Tcal (B_4)$
which both are fibred,
\begin{equation}\label{4.28}
\pi :\quad \Tcal (M^4) \to M^4\ ,
\end{equation}
over $M^4=S^4$ or $B_4$ with $\C P^1=\,$SU(2)/U(1) as a typical fibre.

\smallskip

\noindent
{\bf K\"ahler structure on twistor spaces $\Tcal (S^4)$ and $\Tcal (B_4)$}.
Let us consider a trivial complex vector bundle $\Tcal (M^4){\times}\C^4$
with the flat connection
\begin{equation}\label{4.29}
\hat\Acal =\Qh^{-1}\diff\Qh=G^{-1}\Acal G + G^{-1}\diff G =:
\begin{pmatrix} \Ah^-&-\ve\hat\p\\ \hat\p^\+ & \Ah^+\end{pmatrix}\ ,
\end{equation}
where
\begin{equation}\label{4.30}
\hat\p = \p g=:-\frac{\im}{2\La}\begin{pmatrix}\o^2&\o^{\1}\\
\o^1& -\o^{\2}\end{pmatrix}\ ,\quad
\Ah^-=A^-=\begin{pmatrix}\a_-&-\bb_-\\
\b_-& -\a_-\end{pmatrix}\ ,\quad
\Ah^+=:\begin{pmatrix}\hat\a_+&-\frac{1}{2R}\,\o^{\3}\\
\frac{1}{2R}\,\o^{3}& -\hat\a_+\end{pmatrix}\ ,
\end{equation}
with $\a_-$, $\b_-$ given in (\ref{4.12}) and
\begin{equation}\label{4.31}
\hat\a_+:=\frac{1}{1+\zeta\zeb}\left\{(1-\zeta\zeb)\,\a_+ + \zeb\b_+
-\zeta\bb_+ +\frac{1}{2}(\zeb\diff\zeta - \zeta\diff\zeb )\right\}\ ,
\end{equation}
\begin{equation}\label{4.32}
\o^1:=\frac{1}{(1+\zeta\zeb)^{\frac{1}{2}}}\, (\th^1-\zeta\th^{\2})\ ,
\quad
\o^2:=\frac{1}{(1+\zeta\zeb)^{\frac{1}{2}}}\, (\th^2+\zeta\th^{\1})\ ,
\end{equation}
\begin{equation}\label{4.33}
\o^3:=\frac{2R}{(1+\zeta\zeb)^{\frac{1}{2}}}\, (\diff\zeta +\b_+ -
2\zeta\a_+ +\zeta^2\bb_+)\ .
\end{equation}
Note that forms (\ref{4.32}) and (\ref{4.33}) define on $\Tcal (M^4)$ an
integrable almost complex structure~\cite{AHS} $\J =\J_+$ such that
\begin{equation}\label{4.34}
\J\,\o^i = \im\,\o^i
\end{equation}
with $i=1,2,3$. In other words, $\o^i$'s are (1,0)-forms w.r.t. $\J$.

{}From flatness of the connection (\ref{4.29}), $\diff\hat\Acal +
\hat\Acal\wedge\hat\Acal =0$, we obtain the equations
\begin{equation}\label{4.35}
\diff\begin{pmatrix}\o^1\\ \o^2\\ \o^3\end{pmatrix}=
\begin{pmatrix}\hat\a_++\a_-&-\b_-&\sfrac{1}{2R}\o^{\2}\\
\bb_-&\hat\a_+-\a_-&-\sfrac{1}{2R}\o^{\1}\\
-\sfrac{\ve R}{2\La^2}\,\o^2& \sfrac{\ve R}{2\La^2}\,\o^1& 2\hat\a_+
\end{pmatrix}\wedge\begin{pmatrix}\o^1\\ \o^2\\ \o^3\end{pmatrix}
\end{equation}
defining the connection on $\Tcal (M^4)$ with $M^4=S^4$
or $B_4$. For both cases the metric on $\Tcal (M^4)$ has the form
\begin{equation}\label{4.36}
\diff s^2_{\ve} = \o^1\o^{\1} +\o^2\o^{\2} + \ve\o^3\o^{\3}
\end{equation}
and the almost K\"ahler 2-form $\o$ reads\footnote{For $\ve =-1$
the metric (\ref{4.36}) is not positive definite and one can say
about pseudo-Hermitian metric, pseudo-K\"ahler 2-form etc. but we will
avoid this.}
\begin{equation}\label{4.37}
\o_{\ve} = \sfrac{\im}{2}\,(\o^1\wedge\o^{\1} +\o^2\wedge\o^{\2} +
\ve\o^3\wedge\o^{\3})
\end{equation}
From (\ref{4.35}) one obtains that the 2-form (\ref{4.37}) is K\"ahler,
i.e. $\diff\o_{\ve}=0$, if and only if $R^2=\La^2$. In this case
(\ref{4.35}) defines for $\ve =1$ the Levi-Civita connection with U(3)
holonomy group on $\Tcal (S^4)=\,$Sp(2)/Sp(1)$\times$U(1) $\cong$
SU(4)/U(3) $\cong \C P^3$~\cite{Hitchin}. Similarly, for $\ve =-1$ the structure
equations (\ref{4.35}) define on $\Tcal (B_4)=\,$Sp(1,1)/Sp(1)$\times$U(1)
the Levi-Civita connection for the metric (\ref{4.36}) with the holonomy
group U(2,1).

\bigskip

\section{K\"ahler geometry on twistor spaces of $\C P^2$ and $\C B_2$}

\noindent
{\bf Manifolds $\C P^2$ and $\C B_2$ as coset spaces}. We introduce $\C P^2$
and $\C B_2$ as coset spaces
\begin{equation}\label{5.1}
\C P^2 = {\rm SU(3)/S(U(1)}{\times}{\rm U(2))}\und
\C B_2 = {\rm SU(1,2)/S(U(1)}{\times}{\rm U(2))}
\end{equation}
and denote both of them as $M^4\cong \C P^2$ or $\C B_2$ with local complex
coordinates $y^{\a}$, $\a =1,2$. Consider now the principal bundles
\begin{equation}\label{5.2}
{\rm SU(3)}\to\C P^2
\end{equation}
and
\begin{equation}\label{5.3}
{\rm SU(1,2)}\to\C B_2\ ,
\end{equation}
both having the structure group S(U(1)${\times}$U(2))$\cong$U(1)${\times}$SU(2).
Local sections of the fibrations  (\ref{5.2}) and (\ref{5.3}) are given by
3${\times}$3 matrices
\begin{equation}\label{5.4}
V=\g^{-1}\begin{pmatrix}1&-\ve T^{\+}\\T&W\end{pmatrix}\in
\left\{\begin{array}{l}{\rm SU}(3)\quad {\rm for}\quad \ve =1\\
{\rm SU}(1,2)\quad {\rm for}\quad \ve =-1\end{array}\right.
\end{equation}
where
\begin{equation}\label{5.5}
T:=\begin{pmatrix}\yb^{\2}\\y^1\end{pmatrix} ,\quad
W:=\g\cdot{\bf 1}_2 - \frac{\ve}{\g +1}\, TT^{\+}\und
\g = (1+\ve T^{\+}T)^{\sfrac{1}{2}}=(1+\ve y^{\a}\yb^{\ab})^{\sfrac{1}{2}}>0
\end{equation}
obey
\begin{equation}\label{5.6}
 W^{\+}=W\ ,\quad WT=T\und W^2=\g^2\cdot{\bf 1}_2-\ve TT^{\+}
\end{equation}
and therefore
\begin{equation}\label{5.7}
 V^{\+}\eta\,V=V\,\eta\,V^{\+}=\eta\with \eta = {\rm diag} (1, \ve , \ve )\ .
\end{equation}
Matrices (\ref{5.4}) with $\ve=\pm 1$ are representative elements for cosets
(\ref{5.2}) and (\ref{5.3}) encoding information about their geometry.

\smallskip

\noindent
{\bf (Anti-)self-dual gauge fields on $\C P^2$ and $\C B_2$}. Let us
introduce a flat connection on the trivial vector bundle $M^4{\times}\C^4$
given by the formula
\begin{equation}\label{5.8}
\Acal = V^{-1}\diff V =: \begin{pmatrix}2b&-\sfrac{\ve}{2\La}\, \th^{\+}\\
\sfrac{1}{2\La} \th&B\end{pmatrix}
\end{equation}
with $b\in u(1)$ and $B\in u(2)$  on $M^4\cong \C P^2$ or $\C B_2$, where
from (\ref{4.4}) we obtain
\begin{equation}\label{5.9}
 b=\frac{\ve}{4\g^2}(T^{\+}\diff T - \diff T^{\+} T)\und
B=\frac{1}{\g^2}(W\diff W -T\diff T^{\+}- \frac{\ve}{2}\diff T^{\+}T -
\frac{\ve}{2}\,T^{\+}\diff T )\ ,
\end{equation}
\begin{equation}\label{5.10}
\th=\frac{2\La}{\g^2}\, W\,\diff T= \begin{pmatrix}\th^{\2}\\
\th^1\end{pmatrix}=\frac{2\La}{\g}\,
\begin{pmatrix}\diff\yb^{\2}\\ \diff y^1\end{pmatrix}-
\frac{2\ve\La}{\g^2(\g +1)}\, \begin{pmatrix}\yb^{\2}\\
y^1\end{pmatrix}(\yb^{\1}\diff y^{1} + y^2\diff\yb^{\2})\ .
\end{equation}
Here, $\th^1$ and $\th^2$ are local orthonormal basis of (1,0)-forms on
$\C P^2$ for $\ve =1$ and $\C B_2$ for $\ve =-1$. The real parameter $\La$
characterizes ``size'' of these cosets.

The flatness condition, $\diff\Acal + \Acal\wedge\Acal =0$, leads to
the following component equations
\begin{equation}\label{5.11}
 f^-:=\diff b=\frac{\ve}{8\La^2}\th^{\+}\wedge\th = -\frac{\ve}{8\La^2}
(\th^{1}\wedge\th^{\1}-\th^{2}\wedge\th^{\2})\ ,
\end{equation}
\begin{equation}\label{5.12}
 B:= B^{+}  - b\cdot{\bf 1}_2=\begin{pmatrix}a_+&-\bar b_+\\
b_+&-a_+\end{pmatrix} - b\cdot{\bf 1}_2\ ,
\end{equation}
\begin{equation}\label{5.13}
 F=\diff B + B\wedge B=\frac{\ve}{4\La^2}\,\th\wedge\th^{\+}=
-\frac{\ve}{4\La^2}\,
\begin{pmatrix}\th^2\wedge\th^{\2}&\th^{\1}\wedge\th^{\2}\\
-\th^1\wedge\th^{2}&-\th^{1}\wedge\th^{\1}\end{pmatrix}=:
F^{+}-f^-\cdot{\bf 1}_2\ ,
\end{equation}
where
\begin{equation}\label{5.14}
 F^{+}=\diff B^{+} + B^{+}\wedge B^{+}=-\frac{\ve}{8\La^2}\,
\begin{pmatrix}\th^{1}\wedge\th^{\1}+\th^2\wedge\th^{\2}&2\th^{\1}
\wedge\th^{\2}\\-2\th^1\wedge\th^{2}&-\th^{1}\wedge\th^{\1}
-\th^2\wedge\th^{\2}\end{pmatrix}\ .
\end{equation}
{}From (\ref{5.11}) and (\ref{5.14}) it follows that $*f^-=-f^-$ and
$*F^{+}= F^{+}$, i.e. $b$ is an anti-self-dual $u(1)$-connection on a
complex line bundle over $M^4$ and $B^{+}$ is a self-dual
$su(2)$-connection on a rank-2 complex vector bundle over $M^4\cong\C P^2$
or $\C B_2$. Note that the field $B^{+}-b\cdot{\bf 1}_2$ can be
identified with the $u(2)$-valued Levi-Civita connection on $M^4$
and then the curvature $F^{+}$ and $f^-\cdot{\bf 1}_2$ will be the
self-dual ($su(2)$-valued) and anti-self-dual ($u(1)$-valued) parts
of the Riemannian curvature tensor of the metric $\diff s^2=\th^1\th^{\1}+
\th^2\th^{\2}$ for $\th^{\a}$ given in (\ref{5.10}).

\smallskip

\noindent
{\bf Homogeneous twistor spaces of $\C P^2$ and $\C B_2$}. Twistor spaces
of $\C P^2$ and $\C B_2$ are the following nonsymmetric coset spaces:
\begin{equation}\label{5.15}
\Tcal (\C P^2) = {\rm SU(3)/U(1)}{\times}{\rm U(1)}\und
\Tcal (\C B_2) = {\rm SU(1,2)/U(1)}{\times}{\rm U(1)}\ .
\end{equation}
They can be described via representative matrices similar to the twistor
spaces $\Tcal (S^4)$ and $\Tcal (B_4)$ discussed before. Namely, we
consider again the 2${\times}$2 matrix (\ref{4.20}) and 3${\times}$3
matrices
\begin{equation}\label{5.16}
\hat G=\begin{pmatrix}1&0\\0&g\end{pmatrix}\und
\hat V=V\hat G\in\left\{\begin{array}{l}{\rm SU}(3)\quad {\rm for}\quad
\ve =1\\{\rm SU}(1,2)\quad {\rm for}\quad \ve =-1\end{array}\right.
\end{equation}
for $V$ given in (\ref{5.4}). The matrix $\hat V$ defines a local section
of the bundle
\begin{equation}\label{5.17}
{\rm SU(3)}\to\Tcal(\C P^2)\qquad{\rm or}\qquad{\rm SU(1,2)}\to\Tcal(\C B_2)
\end{equation}
depending on $\ve=\pm 1$. Both fibrations (\ref{5.17}) have the group
U(1)${\times}$U(1) as a typical fibre. Thus, the matrices (\ref{5.16})
represent the twistor coset spaces  (\ref{5.15}). We again have fibrations
(\ref{4.28}) but with $M^4\cong\C P^2$ or $M^4\cong\C B_2$.

\smallskip

\noindent
{\bf K\"ahler structure on twistor spaces $\Tcal (\C P^2)$ and
$\Tcal (\C B_2)$}. Consider a trivial complex vector bundle
$\Tcal (M^4)\times\C^4\to\Tcal (M^4)$ with $M^4\cong\C P^2$ or
$\C B_2$. A flat connection on this bundle is defined
by formula
\begin{equation}\label{5.18}
\hat\Acal = \hat V^{-1}\diff\hat V =\hat G^{-1}\Acal\,\hat G +
\hat G^{-1}\diff\hat G =:
 \begin{pmatrix}2b&-\sfrac{\ve}{2\La}\,\hat \th^{\+}\\
\sfrac{1}{2\La}\hat\th&\hat B\end{pmatrix} \ ,
\end{equation}
where
\begin{equation}\label{5.19}
 \hat\th = g^{\+}\th =\frac{1}{(1+\zeta\zeb)^{\frac{1}{2}}}
\begin{pmatrix}
\th^{\2}+\bar\zeta\th^1\\ \th^1-\zeta\th^{\2}
\end{pmatrix}
=:
\begin{pmatrix}
 \o^{\2}\\ \o^1
\end{pmatrix}\ ,\qquad
\hat\th^{\+}=\th^{\+}g=(\o^2\ \o^{\1})\ ,
\end{equation}
\begin{equation}\label{5.20}
 \hat B = g^{\+}B\,g +g^{\+}\diff\,g=:
\begin{pmatrix}
 \hat a_+& -\sfrac{1}{2R}\o^{\3}\\ \sfrac{1}{2R}\o^{3}&-\hat a_+
\end{pmatrix} -b\cdot{\bf 1}_2
\end{equation}
with
\begin{equation}\label{5.21}
\hat a_+ =\frac{1}{1+\zeta\zeb}\,\left\{ (1-\zeta\zeb)a_+ + \zeb b_+ -
\zeta\bar b_+ + \frac{1}{2}(\zeb\diff\zeta -\zeta\diff\zeb )\right\}\ ,
\end{equation}
\begin{equation}\label{5.22}
\o^3 =\frac{2R}{1+\zeta\zeb}\, \left(\diff\zeta  + b_+ -2\zeta a_+ +
\zeta^2\bar b_+\right)\ ,
\end{equation}
and $b, a_+, b_+$ are given in (\ref{5.9})-(\ref{5.12}).

{}From flatness of $\hat\Acal$ we obtain
\begin{equation}\label{5.23}
\hat\Fcal = \diff\hat \Acal +\hat\Acal\wedge\hat\Acal  =
 \begin{pmatrix}
2\diff b -\frac{\ve}{4\La^2}\,\hat\th^{\+}\wedge\hat\th&
-\frac{\ve}{2\La}\,(\diff\hat\th^{\+}{+}\hat\th^{\+}\wedge
\hat B{-}2\hat\th^{\+}\wedge b)
\\[2mm]
\frac{1}{2\La}\,(\diff\hat\th{+}\hat B\wedge\hat \th{-}2b\wedge\hat\th)&
\hat F^+ -\diff b\cdot{\bf 1}_2 -\frac{\ve}{4\La^2}\,\hat\th\wedge\hat\th^{\+}
\end{pmatrix} =0
\end{equation}
and therefore
\begin{equation}\label{5.24}
 f^-=\diff b = -\frac{\ve}{8\La^2}\,(\o^1\wedge\o^{\1}-\o^2\wedge\o^{\2})=
-\frac{\ve}{8\La^2}\,(\th^1\wedge\th^{\1}-\th^2\wedge\th^{\2})\ ,
\end{equation}
\begin{equation}\label{5.25}
\hat F^+ = -\frac{\ve}{8\La^2}\,
 \begin{pmatrix}
\o^1\wedge\o^{\1}+\o^2\wedge\o^{\2}& 2\o^{\1}\wedge\o^{\2}\\
-2\o^{1}\wedge\o^{2}&-\o^1\wedge\o^{\1}-\o^2\wedge\o^{\2}
 \end{pmatrix}
\end{equation}
along with
\begin{equation}\label{5.26}
\diff\hat\th + (\hat B - 2b\cdot{\bf 1}_2)\wedge\hat\th =0\ .
\end{equation}

The metric and an almost K\"ahler structure on $\Tcal (\C P^2)$
and $\Tcal (\C B_2)$ read
\begin{equation}\label{5.27}
\diff s^2 = \o^1\o^{\1} + \o^2\o^{\2} + \ve\o^3\o^{\3}\und
\o = \frac{\im}{2}(\o^1\wedge\o^{\1} + \o^2\wedge\o^{\2} + \ve\o^3\wedge\o^{\3})\ ,
\end{equation}
where $\o^i$'s are given in (\ref{5.19}) and (\ref{5.22}). From
(\ref{5.23})-(\ref{5.26}) it follows that $\o$ is K\"ahler, i.e. $\diff \o =0$,
iff $R^2=2\La^2$. Furthermore, from (\ref{5.23})-(\ref{5.26}) we obtain the
structure equations
\begin{equation}\label{5.28}
\diff\begin{pmatrix}\o^1\\ \o^2\\ \o^3\end{pmatrix}=
\begin{pmatrix}\hat a_++3b&0&\sfrac{1}{2R}\o^{\2}\\[1mm]
0&\hat a_+-3b&-\sfrac{1}{2R}\o^{\1}\\[1mm]
-\sfrac{\ve}{2R}\,\o^2& \sfrac{\ve }{2R}\,\o^1& 2\hat a_+
\end{pmatrix}\wedge\begin{pmatrix}\o^1\\ \o^2\\ \o^3\end{pmatrix}
\end{equation}
which define the Levi-Civita U(3) connection on $\Tcal (\C P^2)$
and the Levi-Civita U(1,2) connection on $\Tcal (\C B_2)$.

\bigskip

\section{Nearly K\"ahler and nearly Calabi-Yau twistor spaces}

\noindent
{\bf Definitions}. Let us consider an oriented 6-dimensional manifold
$X^6$ with a Riemannian metric $g$ and an almost complex structure
$\J$ (U(3)-structure). We may choose a local orthonormal basis $\{e^a\}$
of $T^*X^6$ with $a=1,...,6$ such that the metric and the fundamental
2-form $\o$ read
\begin{equation}\label{6.1}
\diff s^2 = \de_{ab}e^ae^b\ ,
\end{equation}
\begin{equation}\label{6.2}
\o = e^1\wedge e^2 + e^3\wedge e^4 + e^5\wedge e^6
\end{equation}
and
\begin{equation}\label{6.3}
\J e^1 = -e^2\ ,\quad \J e^3 = -e^4\und \J e^5 = -e^6\ .
\end{equation}
Then forms $\Theta^i$ of type (1,0) w.r.t. $\J$ read
\begin{equation}\label{6.4}
\Theta^1 = e^1+\im e^2\ ,\quad \Theta^2= e^3 +\im e^4\und
\Theta^3= e^5 +\im e^6\ ,
\end{equation}
so that
\begin{equation}\label{6.5}
\J \Theta^i = \im \Theta^i
\end{equation}
and
\begin{equation}\label{6.6}
\diff s^2 = \Theta^1\Theta^{\1} +\Theta^2\Theta^{\2} +\Theta^3\Theta^{\3}
\und
\o = \sfrac{\im}{2}\,(\Theta^1\wedge\Theta^{\1} +\Theta^2\wedge\Theta^{\2}
+\Theta^3\wedge\Theta^{\3})\ .
\end{equation}

We assume that $c_1(X^6)=0$ and introduce a (3,0)-form
\begin{equation}\label{6.7}
 \Omega:=\Theta^1\wedge\Theta^2\wedge\Theta^3={\rm Re}\,\Omega +
\im\, {\rm Im}\,\Omega
= e^{135}{+}e^{425}{+}e^{416}{+}e^{326}+
\im (e^{136}{+}e^{426}{+}e^{145}{+}e^{235})\ .
\end{equation}
So, our manifold $X^6$ has an SU(3) structure defined by nowhere vanishing
forms $\o$ and $\Omega$. Such a manifold is called {\it nearly K\"ahler} if
$\o$ and $\Omega$ satisfy
\begin{equation}\label{6.8}
\diff\o = 3c\,{\rm Im}\,\Omega\und\diff\Omega = 2c\,\o\wedge\o
\end{equation}
with a constant $c\in\R$. A manifold $(X^6, \J , \o , \Omega)$ is called
{\it nearly Calabi-Yau} manifold~\cite{XuPhD} if
\begin{equation}\label{6.9}
 \diff\o =0\und\diff\,{\rm Im}\,\Omega =0\ .
\end{equation}
For more details see~\cite{CS}-\cite{XuPhD}, \cite{Bryant},
\cite{BM}-\cite{But}.

In two previous sections we have described K\"ahler structures on the
twistor spaces $\Tcal (S^4)$, $\Tcal (\C P^2)$, $\Tcal (B_4)$ and
$\Tcal (\C B_2)$ endowed with integrable almost complex structures.
In this section we provide these spaces with never integrable almost
complex structures and introduce on them nearly K\"ahler or nearly
Calabi-Yau structure.

\smallskip

\noindent
{\bf Nearly K\"ahler structure on $\Tcal (S^4)$}. Consider the almost
K\"ahler twistor space $\Tcal (S^4)$ with the complex structure
$\J =\J_+$~\cite{AHS} such that $\J_+\o^i=\im\,\o^i$ with (1,0)-form
$\o^i$ given in (\ref{4.32}), (\ref{4.33}). Let us introduce the forms
\begin{equation}\label{6.10}
 \Theta^1:=\o^1\ ,\quad \Theta^2:=\o^2\und\Theta^3:=\o^{\3}\ ,
\end{equation}
which are forms of type (1,0) w.r.t. an almost complex structure
$\J = \J_-$~\cite{ES}, $\J\,\Theta^i = \im\,\Theta^i$, defined in
(\ref{6.5}). Note that in terms of $\{e^a\}$ we have
\begin{equation}\label{6.11}
 \J_{\pm}e^1 = -e^2\ ,\quad \J_{\pm}e^3 = -e^4\und \J_{\pm}e^5 = \pm e^6\ .
\end{equation}
Here and in the following we consider $\J =\J_-$ which is never integrable
almost complex structure.

{}From (\ref{4.35}) with $\ve =1$ we get
\begin{equation}\label{6.12}
\diff\begin{pmatrix}\Theta^1\\ \Theta^2\\ \Theta^3\end{pmatrix}=
\begin{pmatrix}\hat \a_++\a_-&-\b_-&0\\
\bb_-&\hat\a_+-\a_-&0\\
0& 0& -2\hat\a_+
\end{pmatrix}
\wedge
\begin{pmatrix}\Theta^1\\ \Theta^2\\ \Theta^3\end{pmatrix}
+\frac{1}{2R}
\begin{pmatrix}\Theta^{\2}\wedge\Theta^{\3}\\ \Theta^{\3}\wedge\Theta^{\1}
\\ \frac{2R^2}{\La^2} \Theta^{\1}\wedge\Theta^{\2}\end{pmatrix}\ ,
\end{equation}
where the first term defines the $su(2)\oplus u(1)$ (torsionful) connection
and the last term defines the Nijenhuis tensor (torsion) with components
$N^i_{\jb\kb}$ and their complex conjugate. Namely, we have
\begin{equation}\label{6.13}
 N^1_{\2\3}=N^2_{\3\1}=\frac{1}{2R}\und N^3_{\1\2}=\frac{R}{\La^2}\ .
\end{equation}
{}From (\ref{6.12}) it follows that the manifold $(\Tcal (S^4), \J, \o ,
\Omega)$ is nearly K\"ahler, i.e. $\o$ and $\Omega$ from (\ref{6.6}) and
(\ref{6.7}) satisfy the equations (\ref{6.8}), if $R^2=\sfrac{1}{2}\La^2$
and $c=\sfrac{1}{2R}$. In this case we have $N^3_{\1\2}=\sfrac{1}{2R}$
and therefore the components
\begin{equation}\label{6.14}
N_{\ib\jb\kb}=\de_{\ib l}\,N^l_{\jb\kb}=\sfrac{1}{2R}\,\ve_{\ib\jb\kb}
\und N_{ijk}=\sfrac{1}{2R}\,\ve_{ijk}
\end{equation}
are totally antisymmetric. The connection with torsion
$T =\sfrac{1}{4}\, N$ has holonomy contained in SU(3). Recall that the
(3,0)-form $\Omega$ from (\ref{6.7}) is a nowhere vanishing global section
of the canonical bundle of $\Tcal (S^4)$ which is a trivial bundle since
the first Chern class of $\Tcal (S^4)$ vanishes, $c_1(\Tcal (S^4))=0$.

\smallskip

\noindent
{\bf Nearly K\"ahler structure on $\Tcal (\C P^2)$}. For the manifold
$\Tcal (\C P^2)$ we use the same redefinition (\ref{6.10}) but with $\o^i$
given by (\ref{5.18})-(\ref{5.22}). This endows $\Tcal (\C P^2)$ with
a nonintegrable almost complex structure defined by (\ref{6.3})-(\ref{6.5}).
Then from (\ref{5.23})-(\ref{5.26}) with $\ve =1$ we obtain the structure
equations
\begin{equation}\label{6.15}
\diff\begin{pmatrix}\Theta^1\\ \Theta^2\\ \Theta^3\end{pmatrix}=
\begin{pmatrix}\hat a_++3b&0&0\\
0&\hat a_+-3b&0\\
0& 0& -2\hat a_+
\end{pmatrix}
\wedge
\begin{pmatrix}\Theta^1\\ \Theta^2\\ \Theta^3\end{pmatrix}
+ \frac{1}{2R}
\begin{pmatrix}\Theta^{\2}\wedge\Theta^{\3}\\ \Theta^{\3}\wedge\Theta^{\1}
\\ \frac{R^2}{\La^2} \Theta^{\1}\wedge\Theta^{\2}\end{pmatrix}\ ,
\end{equation}
where the first term defines $u(1)\oplus u(1)$ connection
and the last term defines torsion with $N^1_{\2\3}=N^2_{\3\1}=
\sfrac{1}{2R}$ and $N^3_{\1\2}=\sfrac{R}{2\La^2}$. For $\Tcal (\C P^2)$,
the conditions (\ref{6.8}) for a manifold to be nearly K\"ahler yield
$R^2=\La^2$ that follows from (\ref{6.15}). Furthermore, for
$R^2=\La^2$ one has
\begin{equation}\label{6.16}
N_{ijk}=\sfrac{1}{2R}\,\ve_{ijk}\und
N_{\ib\jb\kb}=\sfrac{1}{2R}\,\ve_{\ib\jb\kb}\ ,
\end{equation}
so that $T=\sfrac{1}{4}\,N$ is a totally antisymmetric torsion.

\smallskip

\noindent
{\bf Nearly Calabi-Yau structure on $\Tcal (B_4)$}. On $\Tcal (B_4)$
we consider the redefinition (\ref{6.10}) with $\o^i$ from (\ref{4.32}),
(\ref{4.33}) and $\a_+, \b_+, \th^1, \th^2$ given by (\ref{4.13}),
(\ref{4.14}) with $\ve =-1$. This redefinition again corresponds to
the choice of the nonintegrable almost complex structure
(\ref{6.3})-(\ref{6.5}) and $c_1(\Tcal (B_4))=0$. Then from (\ref{4.35})
with $\ve =-1$ one obtains the equations
\begin{equation}\label{6.17}
\diff\begin{pmatrix}\Theta^1\\ \Theta^2\\ \Theta^3\end{pmatrix}=
\begin{pmatrix}\hat \a_++\a_-&-\b_-&0\\
\bb_-&\hat\a_+-\a_-&0\\
0& 0& -2\hat\a_+
\end{pmatrix}
\wedge
\begin{pmatrix}\Theta^1\\ \Theta^2\\ \Theta^3\end{pmatrix}
+\frac{1}{2R}
\begin{pmatrix}\Theta^{\2}\wedge\Theta^{\3}\\ \Theta^{\3}\wedge\Theta^{\1}
\\ -\sfrac{2R^2}{\La^2} \Theta^{\1}\wedge\Theta^{\2}\end{pmatrix}\ ,
\end{equation}
with the $u(2)$ torsional connection defined by the first term and the
Nijenhuis tensor $N^i_{\jb\kb}$ defined by the second term. From (\ref{6.17})
one readily derives that $(\o , \Omega)$ on $\Tcal (B_4)$ satisfy the
nearly Calabi-Yau requirements (\ref{6.9}) if and only if $R^2=\La^2$.
Note also that in this case
\begin{equation}\label{6.18}
\diff \o = \sfrac{1}{\La^2R}\,(R^2-\La^2)\,{\rm Im}\,\Omega =0
\qquad{\rm for}\qquad R^2=\La^2\ ,
\end{equation}
\begin{equation}\label{6.19}
\diff\Omega = -\sfrac{1}{2R}\,
(2\Theta^1\wedge\Theta^2\wedge\Theta^{\1}\wedge\Theta^{\2} -
\Theta^1\wedge\Theta^3\wedge\Theta^{\1}\wedge\Theta^{\3}-
\Theta^2\wedge\Theta^3\wedge\Theta^{\2}\wedge\Theta^{\3})\in
\La^{2,2}(\Tcal (B_4))\ ,
\end{equation}
and therefore\footnote{Recall that $\Omega\equiv\Omega^{3,0}$
and $\diff\Omega=(\diff^{1,0}+\diff^{0,1}+\diff^{-1,2}+\diff^{2,-1})\,
\Omega$, where $\diff^{1,0}=\pa$ and $\diff^{0,1}=\bar\pa$. On
nearly K\"ahler manifolds (\ref{6.20}) is also satisfied due to
(\ref{6.8}).}
\begin{equation}\label{6.20}
\bar\pa\Omega =0\ .
\end{equation}

Thus, we again obtain a manifold with vanishing first Chern class
and SU(3) structure. The manifold $\Tcal (B_4)$ has negative scalar
curvature and can, in principle, be used in string compactifications
to the de Sitter space-time~\cite{Lust}. Compact twistor spaces with
negative scalar curvature can be obtained from $\Tcal (B_4)$ via the
quotients of $B_4$ by a discrete isometry group.

\smallskip

\noindent
{\bf Nearly Calabi-Yau space $\Tcal (\C B_2)$}. In this case we
consider the redefinition (\ref{6.10}) with $\o^i$ from (\ref{5.19}),
(\ref{5.22}) and $\th^\a$ given by (\ref{5.10}) with $\ve =-1$.
From (\ref{5.28}) with $\ve =-1$ we obtain the structure equations
\begin{equation}\label{6.21}
\diff\begin{pmatrix}\Theta^1\\ \Theta^2\\ \Theta^3\end{pmatrix}=
\begin{pmatrix}\hat a_++3b&0&0\\
0&\hat a_+-3b&0\\
0& 0& -2\hat a_+
\end{pmatrix}
\wedge
\begin{pmatrix}\Theta^1\\ \Theta^2\\ \Theta^3\end{pmatrix}
+\frac{1}{2R}
\begin{pmatrix}\Theta^{\2}\wedge\Theta^{\3}\\ \Theta^{\3}\wedge\Theta^{\1}
\\ -\frac{R^2}{\La^2} \Theta^{\1}\wedge\Theta^{\2}\end{pmatrix}\ ,
\end{equation}
defining the $u(1)\oplus u(1)$ connection and the Nijenhuis
torsion $N^i_{\jb\kb}$ on $\Tcal (\C B_2)$. From (\ref{6.21}) we obtain
\begin{equation}\label{6.22}
\diff \o = \sfrac{1}{2\La^2R}\,(2\La^2-R^2)\,{\rm Im}\,\Omega
\und \diff\,{\rm Im}\,\Omega =0\ ,
\end{equation}
so that $\Tcal (\C B_2)$ is a nearly Calabi-Yau space iff $R^2=2\La^2$.
Compact analogues of this manifold with an SU(3) structure can be
obtained via quotients of $\C B_2$ by a discrete isometry group.

\bigskip

\section{Hermitian-Yang-Mills gauge fields on twistor spaces of $S^4$,
         $\C P^2$, $B_4$ and $\C B_2$}

We have described K\"ahler, nearly K\"ahler and nearly Calabi-Yau
structures on the twistor spaces $\Tcal (S^4)$, $\Tcal (\C P^2)$,
$\Tcal (B_4)$ and $\Tcal (\C B_2)$. Now we will discuss in more details
some explicit solutions of the Hermitian-Yang-Mills
equations defined on bundles $\hat E$ over these manifolds.

\smallskip

\noindent
{\bf K\"ahler $\Tcal (S^4)$ and $\Tcal (B_4)$}. Let us consider forms
$\o^i$ of type (1,0) w.r.t. $\J =\J_+$ given in (\ref{4.32}), (\ref{4.33}),
the metric (\ref{4.36}) and the (almost) K\"ahler (1,1)-form (\ref{4.37}).
Consider again the flat connection (\ref{4.29}) for which we have
\begin{equation}\label{7.1}
\hat\Fcal = \diff\hat\Acal + \hat\Acal\wedge\hat\Acal = \begin{pmatrix}
\Fh^--\ve\,\hat\p\wedge\hat\p^\+&-\ve (\diff\hat\p + \hat\p\wedge \Ah^+ +
\Ah^-\wedge\hat\p)\\
\diff\hat\p^\+ + \Ah^+\wedge\hat\p^\+ + \hat\p^\+\wedge \Ah^-& \Fh^+-
\ve\,\hat\p^\+\wedge\hat\p\end{pmatrix} =0\ ,
\end{equation}
where $\hat\p$ and $\Ah^\pm$ are given in (\ref{4.30}). From (\ref{7.1})
it follows that
\begin{equation}\nonumber
\Fh^- = \ve\,\hat\p\wedge\hat\p^\+ = - \frac{\ve}{4\La^2}
\begin{pmatrix} \o^1\wedge\o^{\1} - \o^2\wedge\o^{\2}& 2\o^{\1}\wedge\o^{2}\\
 -2\o^1\wedge\o^{\2}&-\o^1\wedge\o^{\1} + \o^2\wedge\o^{\2}
\end{pmatrix}=
\end{equation}
\begin{equation}\label{7.2}
= - \frac{\ve}{4\La^2}\begin{pmatrix}
\th^1\wedge\th^{\1} - \th^2\wedge\th^{\2}& 2\th^{\1}\wedge\th^2\\
-2\th^1\wedge\th^{\2}&-\th^1\wedge\th^{\1} + \th^2\wedge\th^{\2}\end{pmatrix}
=\ve\,\p\wedge\p^\+ =F^-\ ,
\end{equation}
\smallskip
\begin{equation}\nonumber
\Fh^+ = \ve\,\hat\p^\+\wedge\hat\p =  g^\+F^+g= - \frac{\ve}{4\La^2}
\begin{pmatrix} \o^1\wedge\o^{\1} + \o^2\wedge\o^{\2}&
2\o^{\1}\wedge\o^{\2}\\
 -2\o^1\wedge\o^{2}&-\o^1\wedge\o^{\1} - \o^2\wedge\o^{\2}
\end{pmatrix}=
\end{equation}
\smallskip
\begin{equation}\label{7.3}
= {-}\frac{\ve}{2\La^2(1{+}\zeta\zeb)}\begin{pmatrix}
\sfrac{1}{2}(1{-}\zeta\zeb)(\th^{1\1}{+}\th^{2\2}){+}\zeta\th^{\1\2}{-}\zeb\th^{12}&
\th^{\1\2}{-}\zeb(\th^{1\1}{+}\th^{2\2}){+}\zeb^2\th^{12}    \\[2mm]
-[\th^{12}{+}\zeta(\th^{1\1}{+}\th^{2\2}){+}\zeta^2\th^{\1\2}] &
-\sfrac{1}{2}(1{-}\zeta\zeb)(\th^{1\1}{+}\th^{2\2}){-}\zeta\th^{\1\2}{+}\zeb\th^{12}
\end{pmatrix}\ .
\end{equation}
Recall that we use hats for fields on $\Tcal (M^4)$ and denote fields on $M^4$
by letters without hats.

{}From (\ref{7.2}) it follows that
\begin{equation}\label{7.4}
 (\Fh^-)^{0,2}=0\und\o\,\lrc\,\Fh^- =0\ ,
\end{equation}
i.e. the $su(2)$-valued gauge field $ \Fh^-$ satisfies the HYM equations on
$\Tcal (M^4)$ with $M^4=S^4$ or $B_4$. This solution is a pull-back to
$\Tcal (M^4)$ of the ASD gauge field $F=F^-$ on $M^4=S^4$ or $B_4$. However,
there are solutions of the HYM equations on $\Tcal (M^4)$ which are not
lifted from instantons on $M^4$. To give an example, we rewrite the flat
connection (\ref{4.29}) in the form
\begin{equation}\label{7.5}
\hat\Acal =\begin{pmatrix} \Ah&-\ve\hat T_c\\ \hat T & \hat\a_+\end{pmatrix}
\with \hat T=\frac{\im}{2\La}(\o^1, -\!\o^2, -\!\im\o^3)\und
\hat T_c=-\frac{\im}{2\La}
\begin{pmatrix}
\o^{\1}\\-\o^{\2}\\ \im\ve\o^{\3}
\end{pmatrix}\ .
\end{equation}
Then from the flatness condition
\begin{equation}\label{7.6}
\hat\Fcal=\diff\hat\Acal + \hat\Acal\wedge\hat\Acal =
\begin{pmatrix}
\Fh - \ve\hat T_c\wedge\hat T&
-\ve [\diff\hat T_c+(\Ah + \hat\a_+)\wedge\hat T_c]\\
\diff\hat T+\hat T\wedge (\Ah + \hat\a_+)&
-(\diff\hat\a_++\ve\hat T\wedge\hat T_c)
\end{pmatrix} = 0
\end{equation}
it follows that the Yang-Mills field
\begin{equation}\label{7.7}
\Fh=\diff\Ah+\Ah\wedge\Ah = \frac{\ve}{4\La^2}
\begin{pmatrix}
 -\o^{1\1}&\o^{2\1}&\im\o^{3\1}\\
 \o^{1\2}&-\o^{2\2}&-\im\o^{3\2}\\
 -\im\ve\o^{1\3}&\im\ve\o^{2\3}&-\ve\o^{3\3}
\end{pmatrix}
\end{equation}
satisfies the equations
\begin{equation}\label{7.8}
\Fh^{0,2}=0\und\o\,\lrc\,\Fh = -\sfrac{\ve}{4\La^2}\cdot {\bf 1}_3\ .
\end{equation}
Therefore the connection $\At =\Ah{-}\sfrac{1}{3}\,(\tr\Ah)\cdot {\bf 1}_3$
with the curvature $\Ft =\Fh{-}\sfrac{1}{3}\,(\tr\Fh)\cdot {\bf 1}_3$ satisfies
the HYM equations
\begin{equation}\label{7.9}
\Ft^{0,2}=0\und\o\,\lrc\,\Ft = 0\ .
\end{equation}
{}From (\ref{7.7}) one sees that $\Fh$ and $\Ft$ have nonvanishing components
along $\C P^1\hra\Tcal (M^4)$ and hence they cannot be obtained by the pull-back
of an ASD gauge field on $M^4=S^4$ or $B_4$.

\smallskip

\noindent
{\bf K\"ahler $\Tcal (\C P^2)$ and  $\Tcal (\C B_2)$}. In this case from
(\ref{5.18}) with $\hat B=\hat B^+-b\cdot{\bf 1}_2$ and (\ref{5.23}) it
follows that the Abelian gauge potential
\begin{equation}\label{7.10}
\hat B^-:= {\rm diag}(b, b)
\end{equation}
satisfies the HYM equations for $\Fh^-:=\diff\hat B^-$,
\begin{equation}\label{7.11}
(\Fh^-)^{0,2}=0\und\o\,\lrc\,\Fh^- = 0
\end{equation}
since
\begin{equation}\label{7.12}
\diff b=-\sfrac{\ve}{8\La^2}(\o^1\wedge\o^{\1} - \o^2\wedge\o^{\2})\qquad
\Leftrightarrow\qquad
\o\,\lrc\,\diff b=0\ .
\end{equation}

\smallskip

\noindent
{\bf Nearly K\"ahler $\Tcal (S^4)$ and  $\Tcal (\C P^2)$}. Recall that an
SU(3)-structure $(\Tcal (S^4), \o ,\Omega)$ is nearly K\"ahler if
$R^2=\sfrac{1}{2}\La^2$ and an SU(3)-structure $(\Tcal (\C P^2), \o , \Omega)$
is nearly K\"ahler if $R^2=\La^2$. Assuming  this and
substituting (\ref{6.10}) into (\ref{7.2}), we obtain that
\begin{equation}\label{7.13}
\Fh^-= - \frac{1}{4\La^2}\begin{pmatrix}
\Theta^1\wedge\Theta^{\1} - \Theta^2\wedge\Theta^{\2}&
2\Theta^{\1}\wedge\Theta^2\\
-2\Theta^1\wedge\Theta^{\2}&
-\Theta^1\wedge\Theta^{\1} + \Theta^2\wedge\Theta^{\2}\end{pmatrix}
\end{equation}
is a solution of the HYM equations on $\Tcal (S^4)=\,$Sp(2)/Sp(1)${\times}$U(1)
which is essentially the same as (\ref{7.2}). At the same time, the analogue of
3${\times}$3 matrix $\Fh$ from (\ref{7.7}) does not satisfy the HYM equations on
the nearly K\"ahler space  $\Tcal (S^4)$ contrary to the K\"ahler case. On the other
hand, on the nearly K\"ahler space  $\Tcal (\C P^2)$ we have {\it two} canonical
Abelian connections satisfying the HYM equations on $\Tcal (\C P^2)$,
\begin{equation}\label{7.14}
\hat B_1^-={\rm diag}(b, b)\with \diff b = - \frac{1}{8\La^2}
(\Theta^1\wedge\Theta^{\1} - \Theta^2\wedge\Theta^{\2})
\end{equation}
and
\begin{equation}\label{7.15}
\hat B_2^-={\rm diag}(\hat a_+, -\hat a_+)\with \diff \hat a_+ = -
\frac{1}{8\La^2}(\Theta^1\wedge\Theta^{\1} + \Theta^2\wedge\Theta^{\2}
-2\Theta^3\wedge\Theta^{\3})\ ,
\end{equation}
where $b$ and $\hat a_+$ are introduced in (\ref{5.18})-(\ref{5.21}).
Note that the Abelian gauge potential $b$ is pulled back from $\C P^2$
but $\hat a_+$ is not.

\smallskip

\noindent
{\bf Nearly Calabi-Yau $\Tcal (B_4)$ and  $\Tcal (\C B_2)$}. Recall that
forms $\o$ and $\Omega$ define a nearly Calabi-Yau structure on an
almost complex manifold $X^6$ if they obey equations (\ref{6.9}).
For the twistor space $\Tcal (B_4)$ this yields $R^2=\La^2$ and the twistor
space $\Tcal (\C B_2)$ is a nearly Calabi-Yau manifold if $R^2=2\La^2$.
Assuming  this and substituting (\ref{6.10}) into (\ref{7.1}) with
$\ve =-1$, we obtain that the gauge field
\begin{equation}\label{7.16}
\Fh^-= \diff\Ah^-+\Ah^-\wedge\Ah^-=\frac{1}{4\La^2}
\begin{pmatrix}
\Theta^1\wedge\Theta^{\1} - \Theta^2\wedge\Theta^{\2}&
2\Theta^{\1}\wedge\Theta^2\\
-2\Theta^1\wedge\Theta^{\2}&
-\Theta^1\wedge\Theta^{\1} + \Theta^2\wedge\Theta^{\2}
\end{pmatrix}
\end{equation}
satisfies the HYM equations (\ref{7.4}) on $\Tcal (B_4)$. Note that
(\ref{7.16}) differs by sign from (\ref{7.13}).

Similarly, on nearly Calabi-Yau space $\Tcal (\C B_2)$ we have the
Abelian Hermitian-Yang-Mills connection
\begin{equation}\label{7.17}
\hat B^-={\rm diag}(b, b)\with \diff b=\frac{1}{8\La^2}
(\Theta^1\wedge\Theta^{\1} - \Theta^2\wedge\Theta^{\2})
\end{equation}
which is the pull-back of an Abelian anti-self-dual gauge potential
on $\C B_2$.

\smallskip

\noindent
{\bf Lifted ASD gauge fields}. In section 3 we have shown that
anti-self-dual gauge fields $F=F^-$ on any oriented Riemannian
4-manifolds $M^4$ are pulled back to Hermitian-Yang-Mills
gauge fields on the twistor space $\Tcal (M^4)$ of $M^4$ with
an almost complex structure $\J=\J_+$. The same is true for
the twistor spaces $\Tcal (S^4)$, $\Tcal (\C P^2)$,
$\Tcal (B_4)$ and $\Tcal (\C B_2)$ with the never integrable
almost complex structure $\J=\J_-$ since $\Fh =\pi^*F$ has no
components along $\C P^1_x\hra\Tcal (M^4)$. Using $\Theta^i$
from (\ref{6.10}) on all above-mentioned twistor spaces, we obtain
\begin{equation}\label{7.18}
\Fh^+=\pi^*F^+=\sfrac{1}{2}(\Fh_{1\1}+\Fh_{2\2})(\Theta^{1\1} +
\Theta^{2\2}) +\Fh_{12}\Theta^{12}+\Fh_{\1\2}\Theta^{\1\2}\ ,
\end{equation}
\begin{equation}\label{7.19}
\Fh^-=\pi^*F^-=\sfrac{1}{2}(\Fh_{1\1}-\Fh_{2\2})(\Theta^{1\1} -
\Theta^{2\2}) +\Fh_{1\2}\Theta^{1\2}+\Fh_{2\1}\Theta^{2\1}\ ,
\end{equation}
where $\Theta^{1\1}=\Theta^1\wedge\Theta^{\1}$, $\Theta^{12}=
\Theta^1\wedge\Theta^{2}$ etc. Furthermore, for the components
$\Fh_{i\jb}$ we have the same formulae (\ref{3.13}) and (\ref{3.14})
as for the case of an almost complex structure $\J_+$. Thus, any
anti-self-dual gauge field $F=F^-$ on a vector bundle $E$ over
$M^4=S^4$, $\C P^2$, $B_4$ or $\C B_2$ lifted to the twistor space
$(\Tcal (M^4), \J_-)$ satisfies the Hermitian-Yang-Mills equations
on the pulled-back bundle $\hat E=\pi^*E$ over $\Tcal (M^4)$. Some
particular examples of such solutions to the HYM equations on
$\Tcal (M^4)$ were described in this section. A lot of explicit
solutions of the HYM equations on $\Tcal (S^4)$ can be obtained
by lifting multi-instanton solutions on $S^4$. Their moduli space is
known from the ADHM construction~\cite{ADHM}. Note that for $B_4$
families of solutions to the ASDYM equations were described in~\cite{ST}.
These ASD gauge fields are lifted to the HYM gauge fields on nearly
Calabi-Yau space $\Tcal (B_4)$.
Furthermore, all HYM gauge fields on the nearly Calabi-Yau spaces
$\Tcal (M^4)$ are obtainable from ASD gauge fields on $M^4$ lifted to
$\Tcal (M^4)$. This follows from the constraint equation
$\diff\Omega\wedge\Fcal =0$ which along with $\o\lrc\,\Fcal =0$
yields $\Fcal_{3\3}=0$.

\newpage

\section{Twistor action for bosonic and supersymmetric Yang-Mills
         theories}

\noindent
In the previous sections we considered the spaces $M^4=S^4$, $\C P^2$,
$B_4$ and $\C B_2$ with the nearly K\"ahler twistor spaces $\Tcal (S^4)$,
$\Tcal (\C P^2)$ and the nearly Calabi-Yau twistor spaces $\Tcal (B_4)$,
$\Tcal (\C B_2)$. For all these cases $c_1(\Tcal (M^4))=0$ and on
$\Tcal (M^4)$ we have a nonintegrable almost complex structure $\J$,
an almost Hermitian (1,1)-form $\o$ and a (3,0)-form $\Om$ satisfying
(\ref{6.8}) or (\ref{6.9}) and defining an SU(3)-structure on
$\Tcal (M^4)$. Furthermore,
\begin{equation}\label{8.1}
\bar\pa\,\Om :=\diff^{0,1}\Om =0\und\bar\pa\,\o =0
\end{equation}
on the nearly K\"ahler spaces $\Tcal (S^4)$, $\Tcal (\C P^2)$ and
\begin{equation}\label{8.2}
\bar\pa\,\Om  =0\und\diff\,\o =0
\end{equation}
on the nearly Calabi-Yau spaces $\Tcal (B_4)$, $\Tcal (\C B_2)$ and
their compact quotients. The SU(3)-structure on the above-mentioned
twistor spaces allows us to introduce analogues of holomorphic
Chern-Simons (hCS) theory on Calabi-Yau (super)spaces. We briefly
recall the hCS theory.

\smallskip

\noindent
{\bf Holomorphic Chern-Simons theory on Calabi-Yau manifolds}. Let $\Zcal$
($\cong X^6$) be a complex three-dimensional Calabi-Yau manifold, $\Ecal$
a rank $k$ complex vector bundle over $\Zcal$ and $\Acal$ a connection
one-form on $\Ecal$. Consider the action~\cite{Witt1}
\begin{equation}\label{8.3}
S=\int_{\Zcal}\Om\wedge\tr (\Acal^{0,1}\wedge\bar\pa\Acal^{0,1}+
\sfrac{2}{3}\,\Acal^{0,1}\wedge\Acal^{0,1}\wedge\Acal^{0,1})\ ,
\end{equation}
where $\Om$ is a nowhere vanishing holomorphic (3,0)-form on $\Zcal$
and $\Acal^{0,1}$ is the (0,1)-component of the connection one-form
$\Acal$. This action functional was obtained by Witten~\cite{Witt1}
as a full target space action of the open topological $B$-model on a
complex three-dimensional target space, on which the Calabi-Yau restriction
arises from $N=2$ supersymmetry of the corresponding topological sigma
model and an anomaly cancellation condition.

The field equations following from the action functional (\ref{8.3})
read
\begin{equation}\label{8.4}
\Fcal^{0,2}=\bar\pa\Acal^{0,1}+\Acal^{0,1}\wedge\Acal^{0,1}=0\ .
\end{equation}
Thus, the hCS theory (\ref{8.3}), (\ref{8.4}) describes inequivalent
holomorphic structures $\bar\pa_{\Acal}=\bar\pa +\Acal^{0,1}$ on the
bundle $\Ecal\to\Zcal$.

\smallskip

\noindent
{\bf Holomorphic Chern-Simons theory on Calabi-Yau supermanifolds}.
In~\cite{Witt2} it was observed that the Calabi-Yau restriction on
the manifold $\Zcal$ can be relaxed by considering a topological
B-model (twistor string theory) whose target spaces are Calabi-Yau
supermanifolds. For them, fermionic directions also make a contribution
to $c_1(\Zcal )$ yielding more freedom to have an overall vanishing
first Chern class. As a main example of
$\Zcal$, Witten considered the supertwistor space $\Pcal^{3|4}:=
\C P^{3|4}\setminus\C P^{1|4}$ with embedded projective lines
$\C P^1_{x,\th}$ parametrized by the chiral superspace ${\cal R}^{4|8}
\ni (x^{\m}, \th^{\a A})$, where $\m =1,...,4$, $\a = 1,2$, $A=1,...,4$.
Under some assumptions, including triviality of the bundle
$\Ecal\to\Pcal^{3|4}$ after restriction to each\footnote{This
condition is equivalent to vanishing of a part of the curvature $\Fcal =
\diff\Acal + \Acal\wedge\Acal$ having components along subspaces
$\C P^1_{x,\th}\hra\Pcal^{3|4}$. Without this assumption the hCS theory
is not equivalent to the anti-self-dual $\Ncal =4$ SYM theory.}
$\C P^1_{x,\th}\hra\Pcal^{3|4}$, it was shown that hCS theory on the
supertwistor space $\Pcal^{3|4}$ is equivalent to anti-self-dual
$\Ncal =4$ super-Yang-Mills (SYM) theory in four dimensions.\footnote{For
reductions of this model to $d=3$ and $d=2$ see~\cite{PSW, Popov1}.}

As equations of motion for hCS theory on $\Pcal^{3|4}$ and $\C P^{3|4}$
one has (\ref{8.4}) but with $\Acal^{0,1}$ holomorphically depending on
fermionic coordinates. The spectrum of physical states contained in
$\Acal^{0,1}$ is the same as that of $\Ncal =4$ SYM theory but the
interactions of both theories differ. It was also shown that the
perturbative amplitudes of the full $\Ncal =4$ SYM theory are recovered
by adding to the hCS action a nonlocal term interpreted as D-instantons
wrapping holomorphic curves in $\Pcal^{3|4}$. Another option is to construct
an action on the super-ambitwistor space~\cite{Witt2, PoSa, MaSc} but
this was not entirely successful.

\smallskip

\noindent
{\bf Pseudo-holomorphic Chern-Simons theory}. Recall that twistor string
theory establish a connection with $\Ncal =4$ SYM theory in four dimensions
but, contrary to the standard topological string theory on Calabi-Yau
3-folds, lost the connection with superstring theory. For restoring such a
connection one should consider not the superspace $\C P^{3|4}$ but an
ordinary 6-manifold as a target space for twistor strings. In fact, the
complex twistor space $\C P^{3}$ was used for some proposals on a possible
twistor action for nonsupersymmetric $d=4$ Yang-Mills theory~\cite{Mason}.
However, nearly K\"ahler and/or nearly Calabi-Yau twistor spaces
$\Tcal (M^4)$ may be more suitable for this purpose since all these
twistor spaces carry an SU(3) structure defined by forms $\o$ and $\Om$.
Thus, we can consider the action functional (\ref{8.3}) with $\Zcal =
\Tcal (M^4)$ and $M^4=S^4$, $\C P^2$ or $B_4$, $\C B_2$ (or their compact
quotients) for almost complex $\Zcal$ with $c_1(\Zcal )=0$. In this case,
$\Acal^{0,1}$ will be a (0,1)-form w.r.t. the nonintegrable almost
complex structure $\J=\J_-$ introduced in (\ref{6.3})-(\ref{6.5}) and
(\ref{6.10}). The field equations (\ref{8.4}) of this pseudo-holomorphic
Chern-Simons (pshCS) theory describe inequivalent pseudo-holomorphic
structures $\bar\pa_{\Acal}=\bar\pa +\Acal^{0,1}$ on the bundle $\Ecal\to
\Zcal$. In its turn, pshCS theory on the almost complex twistor space
$\Tcal (M^4)$ is equivalent to the (bosonic) anti-self-dual Yang-Mills
theory on $M^4=S^4$, $\C P^2$, $B_4$, $\C B_2$ or $\R^4$. Thus, one may
consider (\ref{8.3}) as a candidate to a twistor action for bosonic
ASDYM theory and consider nearly K\"ahler \& nearly Calabi-Yau twistor
spaces as candidates for a target space for twistor string
theory, which is close to the standard topological string theory.

\smallskip

\noindent
{\bf Action functionals on nearly K\"ahler twistor spaces}. As it was shown
in section 3, for any anti-self-dual gauge field $F$ on $M^4$, its pull-back
$\Fh :=\pi^*F$ to the twistor space\footnote{Recall that $\pi : \Tcal (M^4)
\to M^4$ is the canonical projection.} $\Tcal (M^4)$ satisfies not only
(\ref{8.4}) but also the equation $\o\,\lrc\ \Fh =0$, where $\o$ is an almost
Hermitian (1,1)-form on $\Tcal (M^4)$. Thus, $\Fh$ is a solution of the
Hermitian-Yang-Mills equations on $\Tcal (M^4)$ which are the BPS equations
for Yang-Mills theory in $d=6$.

It is of interest that on nearly K\"ahler manifolds $X^6$ not only (\ref{8.4})
but the full HYM equations can be obtained from the action functional~\cite{XuPhD}
\begin{equation}\label{8.5}
S=\int_{X^6}{\rm Im}\,\Om\wedge\tr (\Acal\wedge\diff\Acal +
\sfrac{2}{3}\,\Acal\wedge\Acal\wedge\Acal)\ ,
\end{equation}
where $\Acal$ is a connection one-form on a complex vector bundle
$\Ecal$ over $X^6$ and $\Om$ is a (3,0)-form on $X^6$. Note that
$\diff\o = 3c\,{\rm Im}\,\Om$ and therefore in (\ref{8.5}) one can use
$\diff\o$ instead of Im$\,\Om$.

The field equations following from (\ref{8.5}) read
\begin{equation}\label{8.6}
{\rm Im}\,\Om \wedge\Fcal=0\ ,
\end{equation}
where $\Fcal = \diff\Acal +\Acal\wedge\Acal$ is the curvature of $\Acal$.
It is easy to show~\cite{XuPhD} that on nearly K\"ahler manifolds from
(\ref{8.6}) it follows
\begin{equation}\label{8.7}
{\rm Re}\,\Om\wedge\Fcal =0\ ,
\end{equation}
and differentiating (\ref{8.7}) we obtain
\begin{equation}\label{8.8}
\o\wedge\o\wedge\Fcal=0\qquad\Leftrightarrow\qquad\o\,\lrc\ \Fcal =0
\end{equation}
after using (\ref{6.8}) and the Yang-Mills Bianchi identities.
In fact, on nearly K\"ahler manifolds eq.(\ref{8.8}) follows from
(\ref{8.4}) due to (\ref{6.8}).

The above observations allow us to propose (\ref{8.5}) as a twistor
action on $X^6\cong\C P^3$ (or SU(3)/U(1)$\times$U(1)) for the bosonic
ASDYM theory on $S^4$ (or $\C P^2$) after assuming, as in hCS and pshCS
theories, that components of $\Fcal$ along $\C P^1_x\hra X^6$ vanish.
Such $\Fcal$ can be identified with the gauge field $\Fh$ pulled back
from $S^4$ (or $\C P^2$) to $X^6$ with the components defined in (\ref{7.18}),
(\ref{7.19}) and (\ref{3.13}), (\ref{3.14}). Furthermore, for the full
$d=4$ (bosonic) Yang-Mills theory one can use the $d=6$ Yang-Mills action
functional
\begin{equation}\label{8.9}
S=-\int_{\Pcal^3}vol_6\ \tr(\Fh_{ab}\Fh_{ab})\ .
\end{equation}
Integrating (\ref{8.9}) over $\C P^1\hra \Pcal^3$, we obtain the standard
Yang-Mills action on $\R^4$ (on $S^4$ for $X^6\cong\C P^3$). This action
functional is a natural part of the low-energy heterotic string theory.
On the other hand, anti-self-dual Yang-Mills theory on $S^4$ and $\C P^2$ is
related with the Hermitian-Yang-Mills model on the twistor spaces $\C P^3$
and SU(3)/U(1)$\times$U(1), respectively, and with heterotic string theory
compactified  on these nearly K\"ahler spaces. It would be of interest to
study open topological string theories (both A and B types) with such target
spaces. According to~\cite{DGNV}, A-model on $\Tcal (M^4)$ can be a holographic
dual to topological M-theory on a $d=7$ $G_2$-manifold naturally associated
with any nearly K\"ahler space $\Tcal (M^4)$~\cite{AW}.

\smallskip

\noindent
{\bf Hermitian-Yang-Mills equations on supermanifolds}. Our observation on
relation between ASDYM theory on $M^4$ and HYM theory on the twistor space
$\Tcal (M^4)$ of $M^4$ can be useful also in $\Ncal =4$ supersymmetric case.
Namely, consider the {\it complex} supertwistor space $\Pcal^{3|4}=
\C P^{3|4}\setminus\C P^{1|4}$~\cite{PoSa} with holomorphic fermionic coordinates
\begin{equation}\label{8.10}
\th^A := \th^{2A} - \zeta\th^{1A}\ ,
\end{equation}
where $\zeta\in U\subset\C P^1$ is a local coordinate on $\C P^1$ and
$\th^{1A}, \th^{2A}$ are Grassmann variables. Introduce local fermionic
(1,0)-forms
\begin{equation}\label{8.11}
\o^A = \frac{1}{(1+\zeta\bar\zeta)^{\frac{1}{2}}}\,(\diff\th^{2A} -
\zeta\,\diff\th^{1A})
\end{equation}
taking values in the Hermitian line bundle $\Lcal_{+1}$ over $\C P^1$
associated with the Hopf bundle (\ref{4.18}) and the corresponding
(0,1)-forms
\begin{equation}\label{8.12}
\o^{\bar A} = \frac{1}{(1+\zeta\bar\zeta)^{\frac{1}{2}}}\,
(\diff\th^{1A} + \bar\zeta\,\diff\th^{2A})=T^{\bar A\bar B}\overline{\o^B}
\quad{\rm{with}}\quad
T^{\1\2}{=}{-}T^{\2\1}{=}T^{\3\4}{=}{-}T^{\4\3}{=}-1
\end{equation}
taking values in the dual line bundle $\Lcal_{-1}\to\C P^1$. Thus,
holomorphic fermionic ``volume form'' $vol_4\o$ takes values in
$\Lcal_{-4}$ and antiholomorphic fermionic ``volume form''
$vol_4\bar\o$ takes values in $\Lcal_{+4}$. We also introduce odd
(local) vector fields
\begin{equation}\label{8.13}
V_A = \frac{1}{(1+\zeta\bar\zeta)^{\frac{1}{2}}}\,\left
(\frac{\pa\ }{\pa\th^{2A}} - \bar\zeta\,\frac{\pa\ }{\pa\th^{1A}}\right )
\und
V_{\bar A} = \frac{1}{(1+\zeta\bar\zeta)^{\frac{1}{2}}}\,\left
(\frac{\pa\ }{{\pa\th^{1A}}^{\sfrac{}{}}} +
\zeta\,\frac{\pa\ }{{\pa\th^{2A}}^{\sfrac{}{}}}\right )
\end{equation}
of type (1,0) and (0,1) which are dual to the forms (\ref{8.11}) and
(\ref{8.12}), respectively.\footnote{Note that one can use a ``nonsymmetric''
formulation by erasing $(1+\zeta\bar\zeta)^{-\sfrac{1}{2}}$ in (\ref{8.13})
and using $(1+\zeta\bar\zeta)^{-1}$ in (\ref{8.11}), (\ref{8.12}). Then
$V_{\bar A}$ will take values in the holomorphic line bundle ${\cal O}(1)
\to\C P^1$, $\o^{\bar A}$ will be a smooth section of the bundle
${\cal O}(-1)$, $V_A\in\bar{\cal O}(1)$ and $\o^{A}\in\bar{\cal O}(-1)$.}
For discussion of reality conditions for odd variables $\th^{\a A}$ and more
details see e.g.~\cite{PoSa}.

Let us consider a holomorphic vector bundle $\Ecal$ over Calabi-Yau
supermanifold $\Pcal^{3|4}$~\cite{Witt2} and a connection one-form
\begin{equation}\label{8.14}
\Acal = \Acal^{\rm b}_i\,\o^i + \Acal^{\rm f}_B\,\o^B +
\Acal^{\rm b}_{\bar i}\,\o^{\bar i}
+ \Acal^{\rm f}_{\bar B}\,\o^{\bar B}=: \Acal^{1,0} + \Acal^{0,1}\ ,
\end{equation}
where $\o^i$ are (1,0)-forms on $\Pcal^3$ (see (\ref{4.32}), (\ref{4.33})
with $\th^1=\diff y$, $\th^2=\diff z$ and $\a_+=\b_+=0$) and by ``b'' and
``f'' we denote even and odd components of $\Acal$. Here, $\Acal^{1,0}$
are given by the first two terms in (\ref{8.14}). On $\Pcal^{3|4}$ we
introduce the (1,1)-form
\begin{equation}\label{8.15}
 \o = \sfrac{\im}{2}\,(\de_{i\bar j}\,\o^i\wedge\o^{\bar j}+
 \de_{A\bar B}\,\o^A\o^{\bar B})\ ,
\end{equation}
where $i,j=1,2,3$ and $A,B=1,...,4$.

The Hermitian-Yang-Mills equations on the supertwistor space $\Pcal^{3|4}$
can be written as follows:
\begin{equation}\label{8.16}
\Fcal^{0,2}=0
\quad\Leftrightarrow\quad
\Fcal_{\bar i\bar j}=0\ ,\quad
\Fcal_{\bar i\bar A}=0\und\Fcal_{\bar A\bar B}=0\ ,
\end{equation}
\begin{equation}\label{8.17}
\o\,\lrc\ \Fcal =0 \quad\Leftrightarrow\quad\de^{i\bar j}\Fcal_{i\bar j}+
\de^{A\bar B}\Fcal_{A\bar B}=0\ ,
\end{equation}
\begin{equation}\label{8.18}
\Fcal^{2,0}=0
\quad\Leftrightarrow\quad
\Fcal_{ij}=0\ ,\quad
\Fcal_{iA}=0\und\Fcal_{AB}=0\ .
\end{equation}
Here,
\begin{equation}\label{8.19}
 \Fcal_{\bar i\bar j}=[V_{\bar i}+\Acal_{\bar i}^{\rm b}, V_{\bar j}+
\Acal_{\bar j}^{\rm b}]\ ,\quad
\Fcal_{\bar i\bar A}=[V_{\bar i}+\Acal_{\bar i}^{\rm b}, V_{\bar A}+
\Acal_{\bar A}^{\rm f}]\ ,\quad
\Fcal_{\bar A\bar B}=\left\{
V_{\bar A}+\Acal_{\bar A}^{\rm f}, V_{\bar B}+
\Acal_{\bar B}^{\rm f}\right\}\ ,
\end{equation}
and similar for other components of $\Fcal$.

\smallskip

\noindent
{\bf Twistor action for $\Ncal =4$ SYM theory}. Let us introduce
\begin{equation}\label{8.20}
 (x^{\a\dot\a})=
\begin{pmatrix}
 x^{1\dot 1}&x^{1\dot 2}\\x^{2\dot 1}&x^{2\dot 2}
\end{pmatrix}=
\begin{pmatrix}
 \zb&\yb\\y&-z
\end{pmatrix}\ ,\quad
(\th^{\a A})= (\th^{\a \dot\a}, \ \th^{\a {\a}'})=
\begin{pmatrix}
 \th^{1\dot 1}&\bar\th^{2\dot 1}&\th^{11'}&\bar\th^{21'}\\
\th^{2\dot 1}&-\bar\th^{1\dot 1}&\th^{21'}&-\bar\th^{11'}
\end{pmatrix}
\end{equation}
and
\begin{equation}\label{8.21}
 (\zeta_\a ):=\rho
\begin{pmatrix}
-\zeta \\1
\end{pmatrix}, \
(\zeta^\a ):=\rho
\begin{pmatrix}
1 \\ \zeta
\end{pmatrix}, \
(\hat\zeta_\a ):=\rho
\begin{pmatrix}
1 \\ \bar\zeta
\end{pmatrix}, \
(\hat\zeta^\a ):=\rho
\begin{pmatrix}
\bar\zeta \\-1
\end{pmatrix}\ {\rm with}\
\rho{:=}\frac{1}{(1+\zeta\bar\zeta)^{\frac{1}{2}}}\ ,
\end{equation}
where in (\ref{8.20}) we used the Euclidean reality conditions~\cite{PoSa}
for $x^{\a\dot\a}$ and $\th^{\a A}$. Using (\ref{8.21}), we can rewrite
(\ref{8.10})-(\ref{8.13}) as
\begin{equation}\label{8.22}
\o^A=\zeta_\a\,\diff\th^{\a A}\ ,\quad
V_A=-\hat\zeta^\a \frac{\pa\ }{\pa\th^{\a A}}\ ,\quad
\o^{\bar A}=\hat\zeta_\a\,\diff\th^{\a A}\and
V_{\bar A}=\zeta^\a \frac{\pa\ }{\pa\th^{\a A}}\ .
\end{equation}

The standard $\Ncal =4$ anti-self-dual Yang-Mills equations~\cite{Vol,
Siegel} can be written in terms of gauge potential components
$\Acal_{\a\dot\a} (x, \th )$ and $\Acal_{\a A} (x, \th )$ and
after introducing
\begin{equation}\label{8.23}
 \bar V_{\dot\a}:=\zeta^{\a}\frac{\pa}{\pa x^{\a\dot\a}}\ ,\quad
\bar \Acal_{\dot\a}^{\rm b}:=\zeta^{\a}\Acal_{\a\dot\a}\ ,\quad
\Acal_{\3}^{\rm b}=0\and
\Acal_{\bar A}^{\rm f}:=\zeta^{\a}\Acal_{\a A}\ ,
\end{equation}
they are equivalent to eqs.(\ref{8.16}) (see e.g.~\cite{PoSa})
with
\begin{equation}\label{8.24}
V_{\1} + \Acal_{\1}^{\rm b} =\bar V_{\dot 2} + \bar\Acal_{\dot 2}^{\rm b}\und
V_{\2} + \Acal_{\2}^{\rm b} =\bar V_{\dot 1} + \bar\Acal_{\dot 1}^{\rm b}
\end{equation}
due to our definition of spinor and vector indices, and eqs.(\ref{8.18})
are Hermitian conjugate to  (\ref{8.16}) for the reality conditions
(\ref{8.20}).
In fact, (\ref{8.23}) defines the pull-back of gauge fields from
${\cal R}^{4|8}$ to $\Pcal^{3|4}$. Moreover, one can show by direct calculations
that (\ref{8.17}) is also equivalent to the $\Ncal =4$ ASDYM equations for this
Hermitian gauge. This is similar to the bosonic case.

Establishing the equivalence of the $\Ncal =4$ ASDYM equations in four
dimensions and the HYM equations (\ref{8.16})-(\ref{8.18}) on the bundle
$\Ecal$ over the supertwistor space $\Pcal^{3|4}$, we can use such advantages
of the twistor description as extended gauge symmetries. Namely, for the
holomorphic bundle $\Ecal\to\Pcal^{3|4}$ one can always find a complex gauge
transformation such that
\begin{equation}\label{8.25}
\Acal^{0,1}-\Acal^{\rm b}_{\3}\o^{\3}\ \to\ \tilde\Acal^{0,1}-
\tilde\Acal^{\rm b}_{\3}\o^{\3}=
g^{-1}(\Acal^{0,1}-\Acal^{\rm b}_{\3}\o^{\3})g +
g^{-1}(\bar\pa-\o^{\3}V_{\3})g=0\ ,
\end{equation}
where $g\in\ $SL$(k,\C)$  and $\Acal^{0,1}-\Acal^{\rm b}_{\3}\o^{\3}$ have components
$\Acal^{\rm b}_{\bar\a}$ and $\Acal^{\rm f}_{\bar A}$.\footnote{By the pull-back
construction $\Acal^{\rm b}_{\3}=0$ but in general $\tilde\Acal^{\rm b}_{\3}\ne 0$.
However, $\tilde\Fcal_{3\3}=g^{-1}\Fcal_{3\3}g=0$.} The equations (\ref{8.16}) in
this gauge dissappear (resolved automatically) since $\tilde\Acal^{\rm b}_{\ab}=0$,
$\tilde\Acal^{\rm f}_{\bar A}=0$, and (\ref{8.17}) reduce to the
equations
\begin{equation}\label{8.26}
\de^{\bar\a\a}V_{\bar\a}\tilde\Acal^{\rm b}_{\a} +
\de^{\bar AA}V_{\bar A}\tilde\Acal^{\rm f}_{A}=0
\end{equation}
which are solved as
\begin{equation}\label{8.27}
\tilde\Acal^{\rm b}_{1}=-V_{\2}\Ups\ ,\quad
\tilde\Acal^{\rm b}_{2}=V_{\1}\Ups\and
\tilde\Acal^{\rm f}_{A}=V_{\bar A}\Ups\ ,
\end{equation}
where the $sl(k, \C)$-valued prepotential $\Ups$ has weight $-2$, i.e.
takes value in the bundle $\Lcal_{-2}$ over $\C P^1$. Substituting (\ref{8.27})
into the rest equations (\ref{8.18}), we obtain three group of equations
(cf.~\cite{Siegel}): one equation without the Grassmann derivatives and two
groups with $V_A$  entering linearly. The equations with the derivatives
$V_A$ simply fix the dependence of $\Ups$ on $\th^A$ in terms of the
``physical'' field
\begin{equation}\label{8.28}
\Phi (x, \zeta , \bar\zeta , \th^{\bar A})=
\Ups(x, \zeta , \bar\zeta , \th^A, \th^{\bar A})\mid_{\th^A=0}
\end{equation}
and its derivatives.

We omit the details here\footnote{The details will be published elsewhere.}
and write out only final formulae. Namely, (\ref{8.18}) reduce to the one
equation
\begin{equation}\label{8.29}
(V_1V_{\bar 1}+V_2V_{\bar 2})\Phi +[V_{\bar 1}\Phi , V_{\bar 2}\Phi ]=0
\end{equation}
on the matrix-valued {\it prepotential} $\Phi$ of weight $-2$ encoding all
information about $\Ncal =4$ ASDYM theory. Expanding the $\Ncal =4$
$sl(k, \C)$-valued prepotential $\Phi$ in $\bar\th^A:=\th^{\bar A}=
\hat\zeta_\a\th^{\a A}$, we
obtain
\begin{equation}\label{8.30}
\Phi = \p_{\a\b}\hat\zeta^{\a}\hat\zeta^{\b}+\p_{\a A}\hat\zeta^{\a}\bar\th^A
+ \p_{AB}\bar\th^A\bar\th^B + \sfrac{1}{3!}\,\chi_{\a}^A\,\zeta^{\a}\,\ve_{ABCD}
\bar\th^A\bar\th^B\bar\th^C + G_{\a\b}\zeta^{\a}\zeta^{\b}\bar\th^1
\bar\th^2\bar\th^3\bar\th^4\ ,
\end{equation}
where ($\p_{AB}(x), \chi_{\a}^A(x), G_{\a\b}(x)$) are space-time fields
of helicities $(0, +\sfrac{1}{2}, +1)$ while $\p_{\a A}(x)$ and
$\p_{\a\b}(x)$ are prepotentials for fields $\tilde\chi_{\dot\a A}$
and $f_{\dot\a\dot\b}$ which have helicities $-\sfrac{1}{2}$ and $-1$.
Finally, the action, whose equations of motion are (\ref{8.29}), have the form
\begin{equation}\label{8.31}
S=\int \diff^4x\,\frac{\diff\zeta\diff\bar\zeta}{(1+\zeta\bar\zeta)^2}\,
vol_4\bar\o\,\tr\left\{\Phi\square\Phi + \frac{2}{3}\,\Phi\,[V_{\bar 1}\Phi ,
V_{\bar 2}\Phi ]\right\} \ ,
\end{equation}
where $\square :=V_1V_{\bar 1}+V_2V_{\bar 2}=\pa_y\pa_{\bar y} +
\pa_z\pa_{\bar z}$. Note that the Lagrangian in (\ref{8.31}) has weight
$-4$ and $vol_4\bar\o$ has weight $+4$ as it should be. The functional
(\ref{8.31}) is the twistor action describing $\Ncal =4$ ASDYM theory in
terms of a single prepotential $\Phi$. Furthermore, one can introduce a
twistor action for the full $\Ncal =4$ SYM theory by adding terms of 2nd,
3rd and 4th degree in $\Phi , \Phi^\+$ and their derivative w.r.t.
$\bar\th^A$ and integrating them with the full Grassmann measure
$vol_4\o\,vol_4\bar\o$.
These terms are
\begin{equation}\nonumber
\Phi^\+\Phi\ ,\quad \bar\th^A\Phi\th^B\Phi^\+\frac{\pa}{\pa\bar\th^A}
\frac{\pa}{\pa\bar\th^B}\Phi\ ,
\end{equation}
\begin{equation}\nonumber
\left(\th^{A_1}\bar\th^{B_1}\frac{\pa}{\pa\bar\th^{A_1}}
\frac{\pa}{\pa\bar\th^{B_1}}\Phi\right)
\left(\th^{A_2}\bar\th^{B_2}\frac{\pa}{\pa\bar\th^{A_2}}
\frac{\pa}{\pa\bar\th^{B_2}}\Phi\right)
\left(\th^{C_1}\th^{D_1}\frac{\pa}{\pa\bar\th^{C_1}}
\frac{\pa}{\pa\bar\th^{D_1}}\Phi\right)
\left(\bar\th^{C_2}\bar\th^{D_2}\frac{\pa}{\pa\bar\th^{C_2}}
\frac{\pa}{\pa\bar\th^{D_2}}\Phi\right)\ .
\end{equation}
Details will be published elsewhere.

\bigskip

\section{Conclusions}

\noindent
In this paper we considered the twistor space $X^6=\Tcal (M^4)$ of an
oriented Riemannian manifold $M^4$ and explored solvability properties
of the first-order Hermitian-Yang-Mills equations for gauge fields on
pseudo-holomorphic bundles $\Ecal$ over $X^6$. It was shown that the
anti-self-dual gauge fields on $M^4$ lifted to $\Tcal (M^4)$ satisfy
the Hermitian-Yang-Mills equations on $\Tcal (M^4)$. Specializing to the
cases $M^4=S^4$, $\C P^2$, $B_4$ or $\C B_2$, we discussed
the nearly K\"ahler and nearly Calabi-Yau structures on their 6-dimensional
twistor spaces $\Tcal (M^4)$ and wrote down some explicit solutions of the
Hermitian-Yang-Mills equations on $\Tcal (M^4)$. Note that for all these
twistor spaces $X^6$ the first Chern class vanishes, $c_1(X^6)=0$, and
these spaces carry an SU(3) structure. We hope that the described
Yang-Mills instanton solutions on the nearly K\"ahler spaces $\Tcal
(S^4)$ and $\Tcal (\C P^2)$ can be used in the flux compactification
of heterotic supergravity to AdS$_4$ and HYM gauge fields on the nearly
Calabi-Yau spaces $\Tcal (B_4)$, $\Tcal (\C B_2)$ and their compact
quotients can be used in the compactifications to the de Sitter space
dS$_4$~\cite{Lust, Toma}. These possibilities will be explored and described
elsewhere.

We have introduced an analogue of holomorphic Chern-Simons theory on nearly
K\"ahler twistor spaces $\Tcal (M^4)$ and shown that under some restrictions
it is equivalent to the anti-self-dual Yang-Mills theory on $M^4=S^4$
or $\C P^2$. A twistor action for non-self-dual Yang-Mills theory is
also proposed. Considering Yang-Mills theory on the supertwistor space
$\C P^{3|4}$ and its open subset ${\cal P}^{3|4}$, we have shown that the
HYM equations encoding the $\Ncal =4$ supersymmetric ASDYM equations reduce
to the equation on a single scalar superfield defined on the supertwistor
space. An expansion of this superfield in Grassmann variables contains all
fields from the $\Ncal =4$ Yang-Mills supermultiplet or prepotentials for
these fields. All terms for a proper twistor action for full $\Ncal =4$ SYM
theory are written down.

A natural direction for further study would be to solve explicitly the
supersymmetry constraint equations of heterotic supergravity by using
solutions to the HYM equations described in this paper. Further study
of twistor actions  for the $\Ncal \le 4$ SYM theories may constitute
another direction. It is also of interest to extend the techniques of the
equivariant dimensional reduction for K\"ahler coset
spaces~\cite{PoSz}-\cite{DoSz} to heterotic supergravity compactified on
six-dimensional nearly K\"ahler and nearly Calabi-Yau coset spaces.

\bigskip

\section*{Acknowledgments}

\noindent
I would like to thank Derek Harland and Olaf Lechtenfeld for useful
remarks. I also wish to thank the Institute for Theoretical
Physics of Hannover University, where this work was completed, for
hospitality. This work was partially supported by the Deutsche
Forschungsgemeinschaft~(grant 436 RUS 113/995) and the Russian Foundation
for Basic Research (grants 08-01-00014-a and 09-02-91347).

\bigskip

\end{document}